\documentclass[aip, amsmath,amssymb,reprint,]{revtex4-1}

\usepackage{graphicx}
\usepackage{dcolumn}
\usepackage{bm}
\usepackage[normalem]{ulem}
\usepackage{xcolor}
\usepackage{adjustbox}
\usepackage{enumitem}
\usepackage{natbib}

\newcommand{\be}{\begin{equation}}
\newcommand{\ee}{\end{equation}}
\newcommand{\ba}{\begin{eqnarray}}
\newcommand{\ea}{\end{eqnarray}}

\newcommand{\red}[1]{\textcolor{red}{#1}}
\newcommand{\blue}[1]{\textcolor{blue}{#1}}

\newcommand{\cyan}[1]{\textcolor{cyan}{#1}}
\newcommand{\green}[1]{\textcolor{green}{#1}}

\newcommand{\darkgreen}[1]{\textcolor{darkgreen}{#1}}
\newcommand{\darkpurple}[1]{\textcolor{darkpurple}{#1}}
\newcommand{\magenta}[1]{\textcolor{magenta}{#1}}

\definecolor{rufous}{rgb}{0.66, 0.11, 0.03}

\newcommand\restr[2]{{
  \left.\kern-\nulldelimiterspace
  #1
  \vphantom{\big|}
  \right|_{#2}
  }}
  
\definecolor{blazeorange}{rgb}{1.0, 0.4, 0.0}
\definecolor{orange}{rgb}{0.902, 0.467, 0.0}
\definecolor{seagreen}{rgb}{0.18, 0.55, 0.34}
\definecolor{darkgreen}{rgb}{0, 0.65, 0}
\definecolor{rufous}{rgb}{0.66, 0.11, 0.03}
\definecolor{royalfuchsia}{rgb}{0.79, 0.17, 0.57}
\definecolor{scarlet}{rgb}{1.0, 0.13, 0.0}
\definecolor{royalpurple}{rgb}{0.47, 0.32, 0.66}
\definecolor{darkblue}{rgb}{0, 0, 0.66}
\definecolor{darkpurple}{rgb}{0.5, 0, 0.5}
\definecolor{lightblue}{rgb}{0.314, 0.522, 0.737}

\usepackage[utf8]{inputenc}
\usepackage[T1]{fontenc}
\usepackage{mathptmx}
\usepackage{etoolbox}

\makeatletter
\def\@email#1#2{
 \endgroup
 \patchcmd{\titleblock@produce}
  {\frontmatter@RRAPformat}
  {\frontmatter@RRAPformat{\produce@RRAP{*#1\href{mailto:#2}{#2}}}\frontmatter@RRAPformat}
  {}{}
}
\makeatother

\begin{document}

\preprint{AIP/123-QED}

\title[Relativistic Shock Reflection using Integral Conservation Laws]{Relativistic Shock Reflection using Integral Conservation Laws}

\author{Jonathan Granot}\email{granot@openu.ac.il.}
 \altaffiliation[Also at ]{Department of Physics, The George Washington University, Washington, DC 20052, USA}
\author{Michael Rabinovich}
\affiliation{Astrophysics Research Center of the Open university (ARCO), The Open University of Israel, P.O Box 808, Ra'anana 4353701, Israel}
\affiliation{Department of Natural Sciences, The Open University of Israel, P.O Box 808, Ra'anana 4353701, Israel}

\date{\today}

\begin{abstract}
Shock wave reflection from a rigid wall has been thoroughly studied in the Newtonian limit, simplifying the problem by analyzing it in a steady-state frame, $S'$, where the point $P$ of the shock's intersection with the wall is at rest. However, a ``super-luminal'' regime emerges when the velocity of point $P$ ($v_p$) exceeds the speed of light ($v_p>c$), where no steady-state frame $S'$ exists. It occurs predominantly in the relativistic regime, relevant in astrophysics, where it encompasses nearly all of the shock incidence angles. To study this regime, we introduce a new approach. We formulate integral conservation laws in the lab frame $S$ (where the unshocked fluid is at rest) for regular reflection (RR), using two methods: \textbf{\textit{a}}. fixed volume analysis and \textbf{\textit{b}}. fixed fluid analysis. We show the equivalence between the two methods, and also to the steady-state oblique shock jump conditions in frame $S'$ in the sub-luminal regime ($v_p<c$). Applying this framework, we find that both the weak and strong shock RR solutions are bounded in the parameter space by the detachment line on the higher incidence angles side. The strong shock solution is also bounded by the luminal line on the lower incidence angles side, and exists only between these two critical lines, in the sub-luminal attachment region.
\end{abstract}

\maketitle

\section{Introduction}
\label{sec:intro}

Supersonic fluid motions can lead to shock waves. At a shock front, there is a sharp increase, which may be approximated as a discontinuity, in the density, pressure and temperature of the incoming upstream fluid, such that they are all larger in the shocked downstream region. In particular, the specific entropy also increases across the shock, which makes the shock an irreversible process. 

In upstream fluid's rest frame, the downstream fluid's velocity just behind the shock is always along the local shock normal. In this frame the shock front velocity along its normal, $v_s$, must be supersonic (exceed the sound speed $c_s$ in the upstream region) for a shock to exist. The ratio of the corresponding proper velocity ($u_s=\Gamma_s\beta_s$ where $\Gamma_s=(1-\beta_s^2)^{-1/2}$, $\beta_s=v_s/c$ and $c$ is the speed of light) to the upstream proper sound speed ($u_{c_s}=\Gamma_{c_s}\beta_{c_s}$ where $\beta_{c_s}=c_s/c$ and $\Gamma_{c_s}=(1-\beta_{c_s}^2)^{-1/2}$) is defined as the shock's Mach number in this frame, $\mathcal{M}=u_s/u_{c_s}$, and $\mathcal{M}>1$ is required for a shock to exist.

When a shock wave encounters an obstacle, such as a rigid body or a wall, the incident shock is reflected off of the obstacle, and a reflected shock is formed. The problem of shock reflection has been extensively studied in the Newtonian regime ($\beta_s\ll 1$), both experimentally and theoretically\citep{Ben-Dor92}. This was pioneered by Ernst Mach\citep{mach1878verlauf}, who reported his discovery in 1878. In
his groundbreaking experimental study (which was later surveyed by\citep{Reichenbach83}) he identified two different
shock wave reflection configurations: (i) a two-shocks configuration that later became known as \textit{regular reflection} (RR), (ii), a three-shocks configuration, that was named after him -- \textit{Mach reflection} (MR). Additional types of \textit{irregular reflection} (IR; a general name for any shock reflection that is not regular reflection) were also discovered later (and are nicely summarized in Ref. \onlinecite{Ben-Dor92}). 

For sufficiently small values of the shock incidence angle (defined as the angle between the incident shock and the wall) only RR is possible, while for sufficiently large incidence angles only MR/IR is possible. In MR, there is a triple point that is detached from the wall, where three shocks intersect: the incident shock, the reflected shock, and a Mach stem that extends between the triple point and the wall \citep{vonNeuman1963,Courant+Friedrichs1948,Chester1954,Hornung1986,Olim+Dewey1992,Rosa+1992,Tabak+Rosales1994}. In RR the incident and reflected shocks meet at a point P on the reflecting wall. Most analytic studies are performed in the rest frame $S'$ of this point, as in this frame the flow is in steady state, which greatly simplifies the analysis. However, for sufficiently small incidence angles and/or fast (relativistic) shocks, the speed of point P in the rest frame $S$ of the unshocked medium exceeds the speed of light, $v_p>c$, and in this super-luminal regime there is no longer a rest frame $S'$ of point P. Therefore, this regime requires an alternative formalism, whose development constitutes a major part of this work.

The super-luminal regime appears mainly in the reflection of relativistic shocks, which in turn are relevant mainly in astrophysics. Shocks are quite common in astrophysical objects and play an important role in many of them, such as in accretion onto the surface of stars, in stellar or galactic winds that interact with their environments, or in cosmological structure formation. Moreover, shocks also appear in astrophysical sources of relativistic jets or outflows, such as pulsar wind nebulae, active galactic nuclei, gamma-ray bursts, micro-quasars, tidal disruption events, or fast radio bursts. Shocks in these sources can form either by collisions between different parts of the outflow (internal shocks) or because of the outflow's interaction with the surrounding medium (external shocks). These shocks accelerate non-thermal relativistic electrons which radiate most of the radiation that we observe from these sources. Moreover, astrophysical shocks may encounter an obstacle and be reflected. Possible examples include the reflection of a gamma-ray burst afterglow shock \citep{Nakar-Granot-07,Lamberts+Daigne2018}, reflection of a collimation shock at a cylindrical jet-cocoon interface \citep{Adamson+Nicholls1958,Norman+1982}, reflection of a supernova shock by a binary companion star \citep{Istomin+Soloviev2008}, or reflection of shock formed at the magnetosphere by the stellar surface of a neutron star or the Sun. Thus motivated, we study the fluid dynamics of shock reflection ranging from the Newtonian to the relativistic regimes, focusing on a planar incident shock colliding at a general incidence angle with a perfectly reflecting wall.
 
In \S\,\ref{sec:reflection} we formulate the relativistic shock reflection problem for RR in the rest frame $S$ of the unshocked fluid, and reduce it to separate treatments and solutions for the incident and reflected shocks.
In \S\,\ref{sec:lab_frame} we formulate the integral conservation laws of particle number, energy and momentum, over either a fixed volume or a fixed fluid.
We start in one dimension
(\S\,\ref{subsec:1d}), showing the equivalence between these two approaches, and between them and the usual steady state shock jump conditions in the rest frame of the shock. Next we generalize the fixed volume (\S\,\ref{sec:2D-fixed-volume}) and fixed fluid (\S\,\ref{sec:2D-fixed-fluid}) methods to two dimensions, showing the equivalence between them.

The solutions of the shock reflection problem are derived in \S\,\ref{sec:solutions}.
We first outline the general
method of solution (\S\,\ref{sec:gen_sol}), and 
proceed with the relatively simpler one 
dimensional case (\S\,\ref{sec:sol_1D}). 
Next, we map the regions within the 
relevant two-dimensional
parameter space (\S\,\ref{sec:sol_2D}) stressing the critical lines (the luminal line, and the detachment line) that divide between regions of one (super-luminal region), two (sub-luminal, attachment region), and no (detachment region) RR solutions. We then present detailed results for the weak shock (\S\,\ref{sec:sol_2DW}) and strong shock (\S\,\ref{sec:sol_2DS}) RR solutions, showing that while both are bounded by the detachment line at large incidence angles, the strong shock solution is also bounded by the luminal line at smaller incidence angles. Finally, we outline and map the dual region (\S\,\ref{sec:dual-region}) where both MR and RR (with either the weak or strong shock solutions) are possible. Our conclusions are discussed in \S\,\ref{sec:dis}. Our analytic results are verified numerically using special-relativistic hydrodynamic simulations in an accompanying paper 
\citep{Bera+23}.

\section{The shock reflection problem: formulation and reduction}
\label{sec:reflection}

\begin{figure}
	\centering
	\includegraphics[trim={0cm 0cm 0cm 0cm},clip,scale=0.6,width=0.48\textwidth]{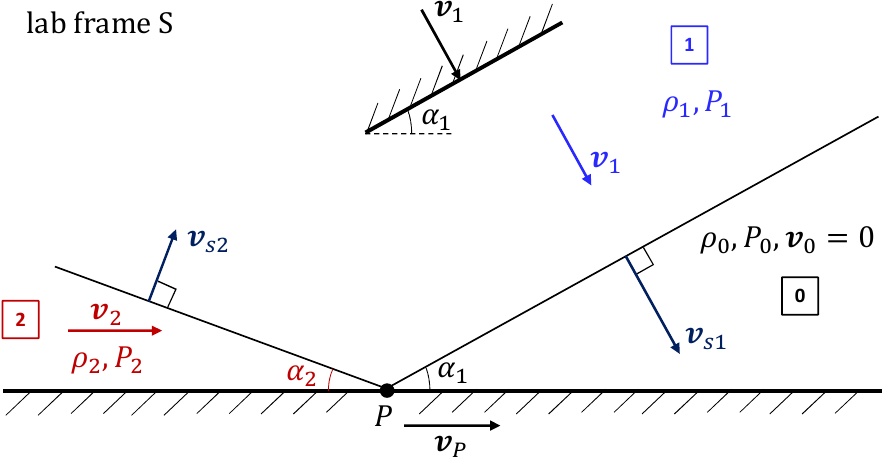}
	\\ \vspace{1.0cm}
	\includegraphics[trim={0cm 0cm 0cm 0cm},clip,scale=0.6,width=0.48\textwidth]{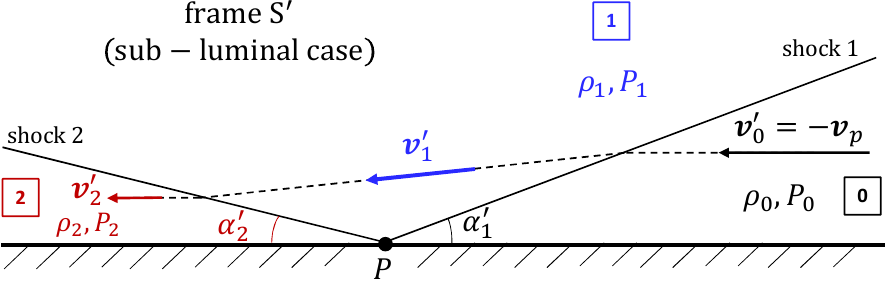}
	\vspace{-0.5cm}
	\caption{\textit{\textbf{Top}}: The shock reflection problem in the rest frame $S$ of the unshocked fluid (region 0). 
\textit{\textbf{Bottom}}:
In the sub-luminal case ($v_p<c$), one can conveniently transform to the frame $S'$ where point $P$ is at rest and the flow is steady and easier to solve.
}
	\label{fig:reflection}
\end{figure}

The well known shock reflection problem may naturally be set up in the rest frame $S$ of the unshocked fluid (region 0), as shown in the \textit{top panel} of Fig.~\ref{fig:reflection}. A piston (ideally distant and infinite) moves at a velocity $\textbf{\textit{v}}_1$ normal to its front (which is at an angle $\alpha_1$ relative to the reflecting wall), which is also the velocity of the singly shocked fluid ahead of it (\textcolor[rgb]{0,0,1.0}{region 1}), behind the shock 1 that it drives into region 0 at a velocity $v_{s1}>v_1$ along its normal. When the incident shock 1 hits the wall, a reflected shock 2 forms that propagates into region 1 at velocity $v_{s2}$ along its normal, its front at an angle $\alpha_2$ relative to the wall, behind which the doubly shocked fluid (\textcolor[rgb]{0.7,0,0}{region 2}) moves parallel to the wall at a speed $v_2$. The instantaneous collision point $P$ moves along the wall at a velocity
\begin{equation}\label{eq:vp-reflection}
v_p = \frac{v_{s1}}{\sin\alpha_1} = \frac{v_{s2}}{\sin\alpha_2}\ .
\end{equation}
In the sub-luminal region ($v_p<c$), one can conveniently transform to the frame $S'$ where point $P$ is at rest, as shown in the \textit{bottom panel} of Fig.~\ref{fig:reflection}. In this frame the flow is steady and therefore easier to solve, e.g. by applying the oblique shock jump conditions there. In the super-luminal region ($v_p>c$) as well as on the luminal line itself ($v_p=c$) no such $S'$ frame exists. This motivates us to develop an alternative method for solving the shock reflection problem, which is formulated in frame $S$ and uses the integral form of the conservation laws.

This problem may be naturally divided into two parts, that can be solved subsequently. First, since shock 1 is unaware of the existence of a reflecting wall until it hits it, then the conditions in region 1 may be readily found by solving the corresponding one-dimensional shock tube problem of a piston driven with velocity $v_1$ into region 0 that is at rest. For the second step, we already know the conditions in region 1 ($p_1$, $\rho_1$, $v_1$) as well as $\textbf{\textit{v}}_{s1}$ (or $v_{s1}$ and $\alpha_1$), and need to solve for the conditions in region 2 ($p_2$, $\rho_2$, $v_2$) as well as $\textbf{\textit{v}}_{s2}$ (or $v_{s2}$ and $\alpha_2$). This second step is addressed in the following section. The main notations used in this work are summarized in Table~\ref{tab:notations}.

\begin{table}
\centering
{\small
\resizebox{0.48\textwidth}{!}{
\begin{tabular}{|c|l|}
\hline
notation & stands for \\
\hline
$P$ & intersection point of the two shocks and the wall\\
$v_p$ & velocity of point $P$ along the wall in the lab frame $S$\\
$S$ & the lab frame, where region 0 and the wall are at rest \\
$S'$ & rest frame of point $P$, where the flow is steady\\
 & (exists only for $v_p<c$) \\
$Q_i$, $Q'_i$
& quantity $Q$ in region $i=0,1,2$ measured in 
frame $S$,\,$S'$
\\
$v_i=\beta_i c$, $v'_i$
& fluid velocity of region $i$ in the rest frame $S$,\,$S'$
\\
$\Gamma_i$, $\Gamma'_i$ 
& Lorentz factor (LF) of region $i$ in rest frame $S$,\,$S'$
\\
$u_i=\Gamma_i\beta_i$, $u'_i$
& proper velocity of region $i$ in rest frame $S$,\,$S'$
\\
$v_{sk}=\beta_{sk} c$ 
& velocity of shock $k=1,2$ along its normal in frame $S$  \\
$\Gamma_{sk}=(1-\beta_{sk}^2)^{-1/2}$ &
the corresponding shock Lorentz factor in frame $S$\\
$u_{sk}=\Gamma_{sk}\beta_{sk}$ &
the corresponding shock proper velocity in frame $S$\\
$\Gamma_{ij}=(1-\beta_{ij}^2)^{-1/2}$ & Lorentz factor of region $i$ relative to region $j$\\
$u_{ij}=\Gamma_{ij}\beta_{ij}$ & relative proper velocity of regions $i$ and $j$\\
$\beta_{sk,i}=-\beta_{i,sk}$
& the velocity of shock $k$ along its normal\\
 & measured in the rest frame of region $i$\\
$\Gamma_{sk,i}$, $u_{sk,i}$ &  the corresponding Lorentz factor and proper velocity\\
$c_{s,i}=\beta_{c_s,i}c$ & sound speed in region $i$ (in the fluid rest frame)\\
$u_{c_s,i}=(\beta_{c_s,i}^{-2}-1)^{-\frac{1}{2}}$ & the corresponding sound proper speed in region $i$\\
$\rho_{i}$ & proper rest-mass density in region $i$\\
$p_i$ & pressure in region $i$ (measured in the fluid rest frame)\\
$e_{{\rm int},i}$ & the proper internal energy density in region $i$ 
\\
$e_i=e_{{\rm int},i}+\rho_i c^2$ & the proper total energy density in region $i$\\
$\hat{\gamma}_i$ & the adiabatic index of the fluid in region $i$\\
$w_i=e_i+p_i$ & the proper  enthalpy density of the fluid in region $i$\\
$h_i=w_i/\rho_i c^2$ & the proper enthalpy per unit rest energy in region $i$\\
\hline
\end{tabular}
}
\caption{Notations used in this work. A quantity (such as density, pressure or time) is said to be ``proper" when it measured in the fluid rest frame.
Proper velocity (or celerity) is the derivative of the observer measured location with respect to the proper time $\tau$, i.e. $\textbf{\textit{u}}_ic=
d\textbf{\textit{x}}_i/d\tau_i=
(dt/d\tau_i)(d\textbf{\textit{x}}_i/dt)=
\Gamma_i\bm{\beta}_ic$.
}
\label{tab:notations}
}
\end{table}

\section{Direct Calculation in the Lab Frame}
\label{sec:lab_frame}

Here we adopt a new approach, which replaces the traditional one of equating the upstream and downstream fluxes (of particle number, energy and momentum components) across the shock in a rest frame (e.g. $S'$) at which the shock front is at rest. Instead, we directly express the conservation laws (of particle number, energy and momentum components) in the lab frame $S$ (in which the unshocked fluid and the ``wall" are at rest) when integrated over a finite volume. We make two different choices for such a reference volume and then show the equivalence of the resulting conservation equations.

\subsection{Three different approaches to a one-dimensional shock}
\label{subsec:1d}

In our setup, a one dimensional shock corresponds to the case $\alpha_1=0$ of a head-on collision with the wall. In this case the incident and reflected shocks do not intersect at a point ($P$) in space, but instead they intersect at a point in time -- the reflection time when the incident shock hits the wall everywhere at once and is instantly replaced by the reflected shock. At this time region 0 disappears and region 2 forms. We consider some later time at which only the reflected shock exists, which separates between regions 1 and 2. In this case $\textbf{\textit{v}}_{2}=0$, i.e. region 2 is at rest in rest frame $S$ -- the original rest frame of region 0, in which we formulate the equations for the integral conservation laws., For this reason $\beta_{s2,2}=\beta_{s2}$, $u_{s2,2}=u_{s2}$ and $\Gamma_{s2,2}=\Gamma_{s2}$.

In such a one dimensional case there is always a frame at which the shock front is at rest and the fluid velocities are normal to it (corresponding to a boost by $\textbf{\textit{v}}_{s2}$ from frame $S$), and it is convenient to use the usual shock jump conditions in this frame. Here we also describe two alternative approaches, which are both equivalent and may be generalized to two (or more) dimensions, where there is not always a frame in which the flow is steady. 

\begin{figure}
	\centering
	\includegraphics[trim={0cm 0cm 0cm 1.0cm},clip,scale=0.6,width=0.48\textwidth]{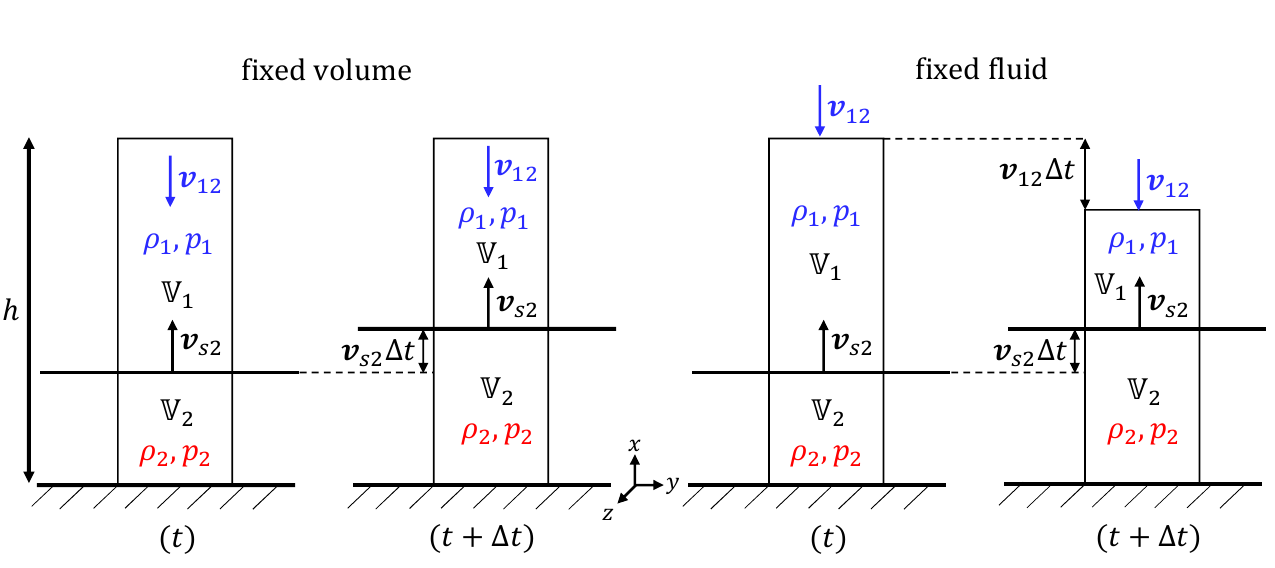}
	\vspace{-0.5cm}
	\caption{\textit{\textbf{Left}}: The fixed volume method, where the total volume $V=V_1+V_2$ is fixed, while the (reflected) shock front that divides regions 1 and 2 moves at a velocity $v_{2s}$ (in the rest frame of region 2 and the wall) such that the volume per unit area on the wall changes as $\dot{V}_2=-\dot{V}_1=v_{2s}$. 
\textit{\textbf{Right}}:
The fixed fluid (or rest mass) method, where a Lagrangian volume is used that follows the same fluid; as the this fluid is shocked it moves from region 1 (that moves at a velocity $v_{12}=-v_{21}$ relative to region 2 and the wall) to region 2 and the respective volumes per unit area on the wall of the fluid within these two regions changes as $\dot{\mathbb{V}}_1 = -(v_{21}+v_{s2})$ and $\dot{\mathbb{V}}_2 = v_{s2}$. 
}
	\label{fig:1Dshock}
\end{figure}

\smallskip
\noindent{\bf Fixed Volume}: Here we consider a fixed volume $V=Ah$ whose base is a unit area ($A=\Delta y\Delta z$) on the wall and height $h$. A quantity $Q$ within this volume (such as mass, energy or momentum component) may simply be obtained by a volume integral over its density $q=dQ/dV$: $Q=\int_V d^3r\, q(\textbf{r}) = V_1q_1+V_2q_2$, where this integral is trivial since regions 1 and 2 are uniform, and occupy volumes $V_1$ and $V_2$, respectively, within the volume $V=V_1+V_2$. Moreover $\dot{V}_2=-\dot{V}_1=Av_{s2}$, where $v_{2s}$ is the velocity of the shock front relative to the fluid in region 2 (and therefore the wall; see \textit{left panels} of Fig.~\ref{fig:1Dshock}). 
The rate of change in $Q$ is simply
\begin{equation}\label{eq:Qdot1}
\dot{Q}=q_1\dot{V}_1+q_2\dot{V}_2 = (q_2-q_1)\dot{V}_2
= (q_2-q_1)Av_{s2}\ .
\end{equation}

On the other hand, $\dot{Q}$ can be calculated by integrating over the flux $\textbf{j}_Q$ of $Q$ into the volume $V$ through its boundaries ($\partial V$), $\dot{Q}=-\int_{\partial V}\textbf{j}_Q\cdot d\textbf{A}$ where $d\textbf{A}=\hat{n}dA$, $dA$ is the differential surface area on the boundary $\partial V$ of $V$, and $\hat{n}$ is the local normal to the surface pointing outward. Since in our case the flow involves two uniform regions then again this integral results in a simple sum, with a contribution from the upper and lower boundaries  in regions 1 and 2, respectively,
\begin{equation}\label{eq:Qdot2}
    \dot{Q} = (j_{Q,2x}-j_{Q,1x})A\ .
\end{equation}
Combining Equations~(\ref{eq:Qdot1}) and (\ref{eq:Qdot2}) gives the general form of the integral conservation equations for a quantity $Q$ over the volume $V$,
\begin{equation}
(q_2-q_1)v_{s2} = j_{Q,2x}-j_{Q,1x}\ .
\end{equation}

\smallskip
\noindent{\bf Fixed Fluid (or rest mass)}:
Here we follow a fixed amount of fluid as part of it is shocked, and for convenience calculate all rates per unit area on the ``wall" (corresponding to $A=\Delta y\Delta z = 1$). As fluid passes through the shock it moves from region 1 to region 2 and the volume it occupies decreases by the shock compression ratio. The rate at which volume containing fluid of region 1 (2) decreases (increases) is given by
\begin{equation}
\dot{\mathbb{V}}_1 = -(v_{12}+v_{s2})\ ,\quad\quad
\dot{\mathbb{V}}_2 = v_{s2}\ ,
\end{equation}
(see \textit{right panels} of Fig.~\ref{fig:1Dshock}). The conservation laws for a quantity $\mathbb{Q}$ (per unit area on the wall) in this case take the form
\begin{equation}
q_1\dot{\mathbb{V}}_1 + q_2\dot{\mathbb{V}}_2 =
q_2v_{s2} - q_1(v_{12}+v_{s2}) =
\dot{\mathbb{Q}}_1+\dot{\mathbb{Q}}_2\ ,
\end{equation}
where $\dot{\mathbb{Q}}_1$ and $\dot{\mathbb{Q}}_2$ are the corresponding source terms from the fixed fluid's interface with regions 1 and 2, respectively. By definition there is no fluid flowing across the Lagrangian boundaries so the source terms vanish in the mass equation ($\dot{\mathbb{Q}}_1=\dot{\mathbb{Q}}_2=0$). For the energy equation the wall is stationary and therefore performs no work ($\dot{\mathbb{Q}}_2=0$), and only the fluid above the upper boundary in region 1 perform work at a rate of $\dot{\mathbb{Q}}_1=p_1v_{21}$ per unit area on the wall. For the momentum as there is no fluid flowing across the boundaries only the pressure contributes to the momentum flux (acting as an external force per unit area) such that $\dot{\mathbb{Q}}_1=-p_1$ and $\dot{\mathbb{Q}}_2=p_2$. 

In Table~\ref{table:1D} we compare the different equations for conservation of mass, energy and momentum using the three different approaches described above. The corresponding sets of equations can be shown to be equivalent, using the relations
\begin{equation}
\Gamma_{s2,1}=\Gamma_{21}\Gamma_{s2}(1+\beta_{21}\beta_{s2})\ ,\quad\quad \beta_{s2,1}=\frac{\beta_{21}+\beta_{s2}}{1+\beta_{21}\beta_{s2}}\ .
\end{equation}

\begin{table*}
    \centering
   \begin{adjustbox}{max width=\textwidth}
    \begin{tabular}{|l|c|c|c|}
    \hline
 conserved & shock jump & \blue{$|q_1|$}\quad fixed \ \ \ \blue{$|j_{Q,1x}|$} & \blue{$|q_1|$}\quad \, fixed \quad\ \ \ \blue{$|\dot{\mathbb{Q}}_1|$}\\
 quantity & conditions & \red{$|q_2|$}\ \  volume\ \ \red{$|j_{Q,2x}|$} & \red{$|q_2|$}\ \ mass/fluid \ \ \red{$|\dot{\mathbb{Q}}_2|$}\\
\hline
mass & $\rho_1 u_{s2,1} = \rho_2 u_{s2}$ & $(\red{\rho_2}-\blue{\rho_1\Gamma_{21}})\beta_{s2} = \blue{\rho_1u_{21}}$ &
$\red{\rho_2}\beta_{s2} - \blue{\rho_1\Gamma_{21}}(\beta_{21}+\beta_{s2}) = 0$\\
energy & $w_1\Gamma_{s2,1}u_{s2,1} = w_2\Gamma_{s2}u_{s2}$ & 
$[\red{e_2}-(\blue{w_1\Gamma_{21}^2-p_1})]\beta_{s2} = \blue{w_1\Gamma_{21}u_{21}}$ &
$\red{e_2}\beta_{s2}-(\blue{w_1\Gamma_{21}^2-p_1})(\beta_{21}+\beta_{s2})=\blue{p_1\beta_{21}}$\\
momentum & $w_1u_{s2,1}^2+p_1 = w_2u_{s2}^2+p_2$ & 
$\blue{w_1\Gamma_{21}u_{21}}\beta_{s2} = \red{p_2}-(\blue{p_1+w_1u_{21}^2})$ &
$\blue{w_1\Gamma_{21}u_{21}}(\beta_{21}+\beta_{s2})=\red{p_2}-\blue{p_1}$\\
\hline
    \end{tabular}
    \end{adjustbox}
    \caption{Comparison of the conservation equations for a one-dimensional shock derived in three different ways.}
    \label{table:1D}
\end{table*}

\subsection{Fixed Volume Method in 2D}
\label{sec:2D-fixed-volume}

Here we generalize the fixed volume method described above to two dimensions, applying it to the second step of the shock reflection problem as formulated in \S\ref{sec:reflection}. We work in frame $S$, which is the rest frame of region 0. The velocities in regions 0 (unshocked fluid), 1 (singly shocked fluid) and 2 (doubly shocked fluid) are
\begin{equation}
    \textbf{\textit{v}}_0=0\,,\quad \textbf{\textit{v}}_1=\beta_1 c(-\cos\alpha_1,\sin\alpha_1)\,, \quad  \textbf{\textit{v}}_2=\beta_2 c(0,1)\ .
\end{equation}
The velocity of shocks 1, $\textbf{\textit{v}}_{s1}$, and 2, $\textbf{\textit{v}}_{s2}$, are defined to be along the respective shock normal,
\begin{equation}
 \textbf{\textit{v}}_{s1}=\beta_{s1} c(-\cos\alpha_1,\sin\alpha_1)\,, \quad  \textbf{\textit{v}}_{s2}=\beta_{s2}c(\cos\alpha_2,\sin\alpha_2)\ .
\end{equation}

\begin{figure}
	\centering
 \includegraphics[trim={0.5cm 0cm 0cm 0cm},clip,scale=0.7,width=0.48\textwidth]{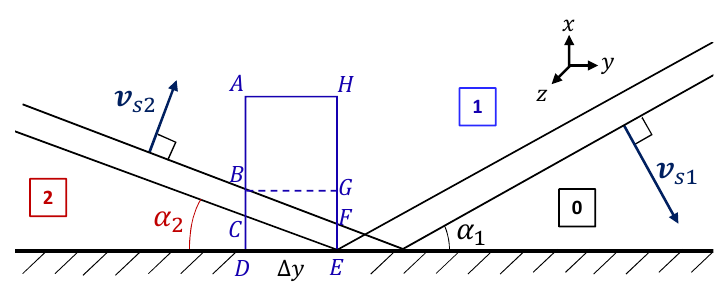}
	\vspace{-0.5cm}
	\caption{
 The volume (in \blue{blue}) where the conservation equations are applied in the \textbf{fixed volume} method}.
	\label{fig:fixed-V}
\end{figure}

We consider a fixed reference volume ADEH containing shock 2, as shown in Fig.~\ref{fig:fixed-V}, whose base is $\Delta y\Delta z$ and height is $\Delta x = {\rm DA} = {\rm EH}$. This volume $V=\Delta x\Delta y\Delta z$ can be divided into the two sub-volumes $V_1$ and $V_2$ containing the fluids in regions 1 and 2, respectively, such that $V=V_1(t)+V_2(t)$ and $\dot{V}_2=-\dot{V}_1=\Delta y\Delta z\frac{v_{s2}}{\cos\alpha_2}$. A quantity $Q$ within this volume (such as mass, energy or momentum component) may simply be obtained by a volume integral over its density $q=dQ/dV$: $Q=\int_V d^3r\, q(\textbf{r}) = V_1q_1+V_2q_2$, where this integral is trivial since regions 1 and 2 are uniform. 
The rate of change in $Q$ is simply

\begin{equation}\label{eq:Qdot1-vol}
\dot{Q}=q_1\dot{V}_1+q_2\dot{V}_2 = (q_2-q_1)\dot{V}_2
= (q_2-q_1)\Delta y\Delta z\frac{v_{s2}}{\cos\alpha_2}\ .
\end{equation}

On the other hand $\dot{Q}$ can be calculated by integrating over the flux $\textbf{j}_Q$ of $Q$ into the volume $V$ through its boundaries ($\partial V$), $\dot{Q}=-\int_{\partial V}\textbf{j}_Q\cdot d\textbf{A}$ where $d\textbf{A}=\hat{n}dA$, $dA$ is the differential surface area on the boundary $\partial V$ of $V$, and $\hat{n}$ is the local normal to the surface pointing outward. Since in our case the flow involves two uniform regions then again this integral results in a simple sum,
\begin{equation}\label{eq:Qdot2-vol}
\begin{aligned}
    \dot{Q}&=\left[(j_{Q,2y}-j_{Q,1y})\cdot GF+j_{Q,2x}\cdot DE - j_{Q,1x}\cdot AH\right]\Delta z\\&=
    [(j_{Q,2y}-j_{Q,1y})\tan\alpha_2+j_{Q,2x}-j_{Q,1x}]\Delta y \Delta z\ .
    \end{aligned}
\end{equation}
Combining Equationss~(\ref{eq:Qdot1-vol}) and (\ref{eq:Qdot2-vol}) gives the general form of the integral conservation equations for a quantity $Q$ over the volume $V$,
\begin{equation}
(q_2-q_1)\frac{v_{s2}}{\cos\alpha_2} = (j_{Q,2y}-j_{Q,1y})\tan\alpha_2+j_{Q,2x}-j_{Q,1x}\ .
\end{equation}
For the conservation of energy and momentum component the relevant densities and fluxes are given by the appropriate components of the energy-momentum tensor, which in our case are
\begin{eqnarray}
\bar{\bar{T}}_1 = \hspace{0.435\textwidth}
\\ \nonumber
\small
\begin{pmatrix}
w_1\Gamma_1^2-p_1 & -w_1\Gamma_1u_1\cos\alpha_1 & w_1\Gamma_1u_1\sin\alpha_1 & 0\\
-w_1\Gamma_1u_1\cos\alpha_1 & w_1u_1^2\cos^2\!\alpha_1+p_1 & -w_1u_1^2\cos\!\alpha_1\sin\!\alpha_1 & 0\\
w_1\Gamma_1u_1\sin\alpha_1 & -w_1u_1^2\cos\!\alpha_1\sin\!\alpha_1 & w_1u_1^2\sin^2\alpha_1+p_1 & 0\\
0 & 0 & 0 & p_1
\end{pmatrix}\,,\quad\  
\hspace{-0.4cm}
\end{eqnarray}
\begin{equation}
\bar{\bar{T}}_2 = 
\begin{pmatrix}
w_2\Gamma_2^2-p_2 & 0 & w_2\Gamma_2u_2 & 0\\
0 & p_2 & 0 & 0\\
w_2\Gamma_2u_2 & 0 & w_2u_2^2+p_2 & 0\\
0 & 0 & 0 & p_2
\end{pmatrix}\ .
\end{equation}
The equations for the continuity of mass, energy as well as $x$ and $y$ components of the momentum, respectively, read 
\begin{eqnarray}\label{eq:SR-M1}
(\rho_2\Gamma_2-\rho_1\Gamma_1)\frac{\beta_{s2}}{\cos\!\alpha_2}=\hspace{0.29\textwidth}\\ \nonumber
(\rho_2u_2-
\rho_1\!\,u_1\!\sin\!\alpha_1)\tan\!\alpha_2 + \rho_1u_1\cos\!\alpha_1\;,
\end{eqnarray}
\begin{eqnarray}\label{eq:SR-E1}
(w_2\Gamma_2^2-p_2-w_1\Gamma_1^2+p_1)\frac{\beta_{s2}}{\cos\alpha_2}=\hspace{0.20\textwidth}\\ \nonumber
(w_2\Gamma_2u_2-w_1\Gamma_1u_1\sin\alpha_1)\tan\alpha_2+w_1\Gamma_1u_1\cos\alpha_1\;,
\end{eqnarray}
\begin{eqnarray}\label{eq:SR-Px1}
w_1\Gamma_1u_1\cos\alpha_1\frac{\beta_{s2}}{\cos\alpha_2}=\hspace{0.28\textwidth}\\ \nonumber
w_1u_1^2\sin\alpha_1\cos\alpha_1\tan\alpha_2+p_2-w_1u_1^2\cos^2\alpha_1-p_1\;,
\end{eqnarray}
\begin{eqnarray}\label{eq:SR-Py1}
(w_2\Gamma_2u_2-w_1\Gamma_1u_1\sin\alpha_1)\frac{\beta_{s2}}{\cos\alpha_2}=\hspace{0.20\textwidth}\\ \nonumber
\quad(w_2u_2^2+p_2-w_1u_1^2\sin^2\!\alpha_1-p_1)\tan\alpha_2+w_1u_1^2\sin\alpha_1\cos\alpha_1\,.
\end{eqnarray}

\subsection{Fixed Fluid (or Rest Mass) Method in 2D}
\label{sec:2D-fixed-fluid}

\begin{figure}
\centering
	\includegraphics[trim={6.5cm 6.5cm 1cm 6.1cm},clip,scale=1,width=0.585\textwidth]{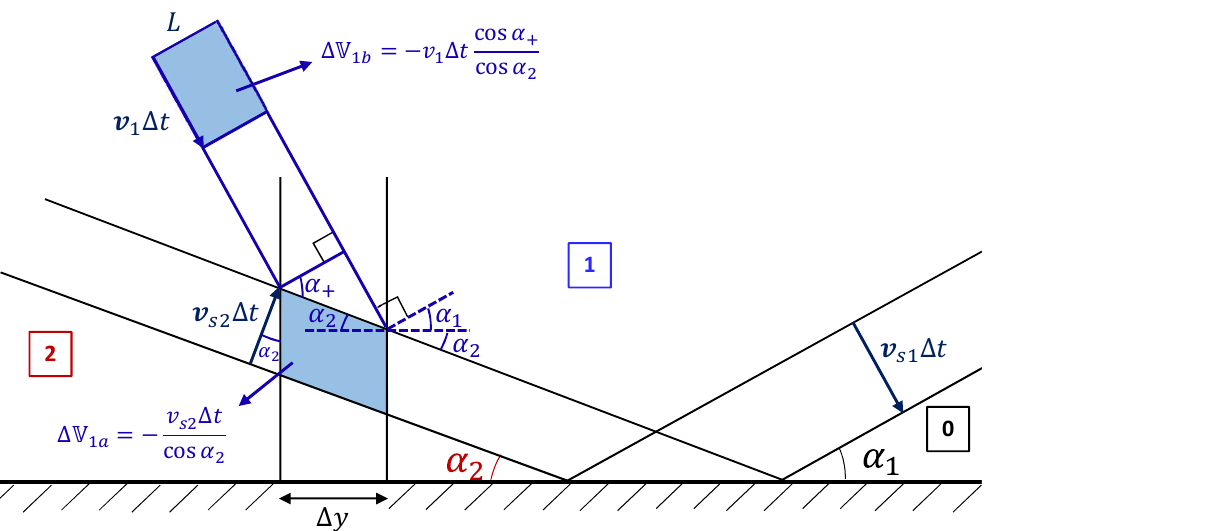}
	\\ \vspace{1.0cm}
	\includegraphics[trim={6cm 6cm 1cm 8cm},clip,scale=1,width=0.58\textwidth]{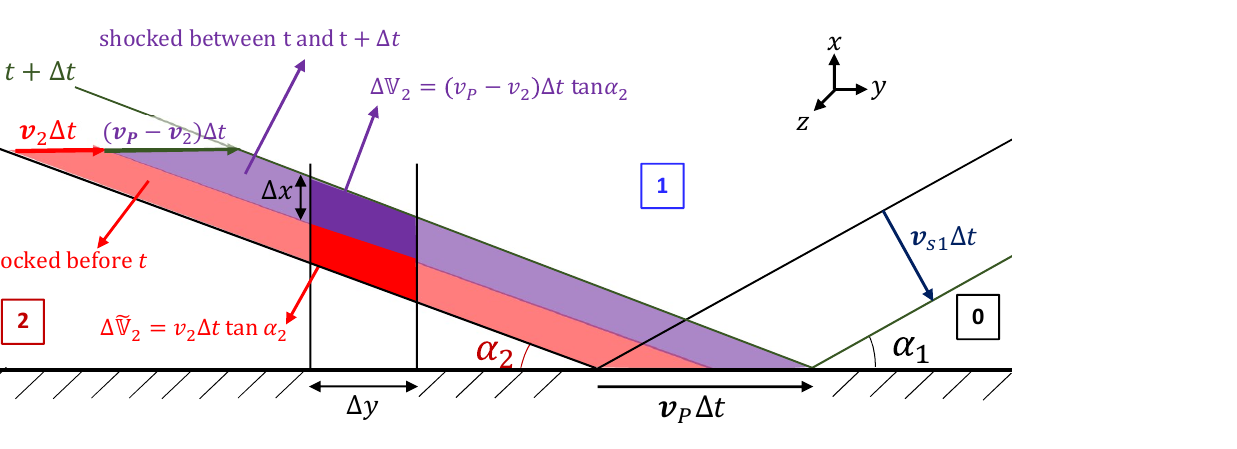}
	\vspace{-0.5cm}
	\caption{\label{fig:fixed_fluid}
 The \textbf{fixed fluid} method: the change over a small time interval $\Delta t$, in the volume per unit area $A=\Delta y\Delta z = 1$ on the wall, within regions 1 ($\Delta\mathbb{V}_1$) and 2 ($\Delta\mathbb{V}_2$) occupied by a fixed amount of fluid.
 \textit{\textbf{Top}}: for region 1 there are two contributions, $\Delta\mathbb{V}_{1a}=-\frac{\Delta z}{A}\Delta y\frac{v_{s2}\Delta t}{\cos\alpha_2}=-\frac{v_{s2}\Delta t}{\cos\alpha_2}$ from the change in the shock front's location, and $\Delta\mathbb{V}_{1b}=-\frac{\Delta z}{A}\Delta y v_{1}\Delta t \frac{\cos\alpha_{+}}{\cos\alpha_2}=-v_{1}\Delta t \frac{\cos\alpha_{+}}{\cos\alpha_2}$ from the flow across the shock at a fixed location. Only the latter contributes to the $pdV$ work performed on the fixed fluid under consideration by the surrounding fluid in region 1 (see Eq.~(\ref{eq:Edot_fixed_fluid})).
\textit{\textbf{Bottom}}: The purple volume $\Delta\mathbb{V}_2=\frac{\Delta z}{A}\Delta y(v_p-v_2)\Delta t\tan\alpha_2=(v_p-v_2)\Delta t\tan\alpha_2$ is added to region 2 between times $t$ and $t+\Delta t$, while the red volume $\Delta\widetilde{\mathbb{V}}_2=v_2\Delta t\tan\alpha_2$ contributes to the $pdV$ work performed on the fixed fluid under consideration by the surrounding fluid in region 2 (see Eq.~(\ref{eq:Edot_fixed_fluid})).
}
\end{figure}

Here we follow a fixed amount of fluid as part of it is shocked, and for convenience calculate all rates per unit area on the ``wall" (corresponding to $A=\Delta y\Delta z = 1$). As fluid passes through the shock it moves from region 1 to region 2 and the volume it occupies decreases by the shock compression ratio. Conveniently denoting $\alpha_+\equiv\alpha_1+\alpha_2$,
for the shock reflection problem the corresponding equations read (see Fig.~\ref{fig:fixed_fluid}),
\begin{eqnarray}\nonumber
\dot{\mathbb{V}}_1 &=& -\frac{v_{s2}+v_1\cos\alpha_+}{\cos\alpha_2} = -\frac{1}{\cos\alpha_2}\left(v_{s1}\frac{\sin\alpha_2}{\sin\alpha_1}+v_1\cos\alpha_+\right)\,,\\
\dot{\mathbb{V}}_2 &=& \left(\frac{v_{s2}}{\sin\alpha_2}-v_2\right)\tan\alpha_2 = \left(\frac{v_{s1}}{\sin\alpha_1}-v_2\right)\tan\alpha_2\ .
\end{eqnarray}
The conservation laws for a quantity $\mathbb{Q}$ (per unit area on the wall) in this case take the form
\begin{equation}
q_1\dot{\mathbb{V}}_1 + q_2\dot{\mathbb{V}}_2 = \dot{\mathbb{Q}}_1+\dot{\mathbb{Q}}_2\ ,
\end{equation}
where $\dot{\mathbb{Q}}_1$ and $\dot{\mathbb{Q}}_2$ are the corresponding source terms from the fixed fluid's interface with regions 1 and 2, respectively. For mass conservation, by construction no fluid passes through the boundary of our Lagrangian volume, so the source terms vanish, leading to
\begin{equation}\label{eq:SR-M2}
\rho_1\Gamma_1\dot{\mathbb{V}}_1\!+ 
\rho_2\Gamma_2\dot{\mathbb{V}}_2 = 0\;\Leftrightarrow\;\frac{\Gamma_2\rho_2}{\Gamma_1\rho_1} = \frac{|\dot{\mathbb{V}}_1|}{\dot{\mathbb{V}}_2}=\frac{1\!+\!\frac{\beta_1}{\beta_{s1}}\frac{\sin\alpha_1}{\sin\alpha_2}\!\cos\alpha_+}{1-\frac{\beta_2}{\beta_{s1}}\sin\alpha_1}\,,
\end{equation}
which is equivalent to Eq.~(\ref{eq:SR-M1}).
For energy conservation the source terms represent the work performed on our fixed fluid at its boundaries with regions 1 and 2, respectively,
\begin{equation}\label{eq:Edot_fixed_fluid}
\dot{\mathbb{E}}_1 = p_1\frac{\Delta\mathbb{V}_{1b}}{\Delta t} = p_1v_1\frac{\cos\alpha_+}{\cos\alpha_2}\ ,\quad\ 
\dot{\mathbb{E}}_2 = p_2\frac{\Delta\tilde{\mathbb{V}}_2}{\Delta t} = p_2v_2\tan\alpha_2\ , 
\end{equation}
and energy conservation reads (upon division by $\tan\alpha_2$)
\begin{eqnarray}\label{eq:SR-E2}
p_1v_1\frac{\cos\alpha_+}{\sin\alpha_2} +  p_2v_2 =
(w_2\Gamma_2^2 - p_2)\left(\frac{v_{s1}}{\sin\alpha_1} - v_2\right)\quad\quad\quad\quad\ \ 
\\ \nonumber
- (w_1\Gamma_1^2-p_1)\frac{v_{s1}}{\sin\alpha_2}\left(\frac{\sin\alpha_2}{\sin\alpha_1}+\frac{\beta_{1}}{\beta_{s1}}\cos\alpha_+\right)\,,
\end{eqnarray}
which can be shown to be equivalent to Eq.~(\ref{eq:SR-E1}).
For the momentum components conservation the source terms represent the respective components of the external force from the pressure at the boundaries with regions 1 and 2, respectively,
\begin{eqnarray}
\dot{\mathbb{P}}_{x1} &=& -p_1\ , 
\quad\quad\quad\quad\ \ 
\dot{\mathbb{P}}_{x2} = p_2\ ,
\\ \nonumber
\dot{\mathbb{P}}_{y1} &=& -p_1\tan\alpha_2\ ,\quad \dot{\mathbb{P}}_{y2} = p_2\tan\alpha_2\ .
\end{eqnarray}
The $x$ and $y$ momentum conservation equations read
\begin{eqnarray}\nonumber
p_2-p_1 &=& 
\frac{\beta_{s1}}{\cos\alpha_2}\left(\frac{\sin\alpha_2}{\sin\alpha_1}+\frac{\beta_1}{\beta_{s1}}\cos\alpha_+\right)w_1\Gamma_1u_1\cos\alpha_1
\\ \label{eq:SR-Px2}
&=& w_1u_1^2\cos^2\alpha_1\left[1+\frac{\tan\alpha_2}{\tan\alpha_1}\left(1+\frac{\beta_{s1}-\beta_1}{\beta_1\cos^2\alpha_1}\right)\right]
\,,\quad\ \ \ 
\end{eqnarray}
\begin{eqnarray}\nonumber
(p_2-p_1)\tan\alpha_2 = w_2\Gamma_2u_2\left(\frac{\beta_{s1}}{\sin\alpha_1}\!-\!\beta_2\right)\tan\alpha_2\hspace{0.1\textwidth}
\\ \label{eq:SR-Py2}
- w_1u_1^2\frac{\sin\alpha_1}{\cos\alpha_2}\left(\frac{\beta_{s1}}{\beta_1}\frac{\sin\alpha_2}{\sin\alpha_1}+\cos\alpha_+\right)\,,\quad\ \ \ 
\end{eqnarray}
which are equivalent to Eqs.~(\ref{eq:SR-Px1}) and (\ref{eq:SR-Py1}), respectively.

\section{Solution of the shock reflection problem}
\label{sec:solutions}

\subsection{The general method of solution}
\label{sec:gen_sol}
The initial free parameters of the problem are $\rho_0$, $p_0$, $e_0$, $v_1$ and $\alpha_1$. If the piston forms a shock, which we label as shock 1, then the shock front is parallel to the piston and its velocity along the shock normal always exceeds that of the piston, $v_{s1}>v_1$. The condition for the formation of a shock is $\mathcal{M}_1=u_{s1}/u_{c_s,0}>1$, and it is always satisfied if $v_1>c_{s,0}$, i.e. when the piston is supersonic. The velocity of shock 1, 
$v_{s1}$, along with the conditions in region 1 ($\rho_1$, $p_1$, $e_1$) may be obtained by solving the 1D shock tube problem, i.e. the shock jump conditions (given in Table~\ref{table:1D}) along with the equation of state in region 1, where in region $i$ (where $i=0,\,1,\,2$),
\begin{equation}\label{eq:EoS-gamma}
p_i = (\hat{\gamma}_i-1)e_{\rm int,i} = (\hat{\gamma}_i-1)(e_i-\rho_ic^2)\ .
\end{equation}
Denoting $\Theta\equiv p/\rho c^2$, $\Theta_i=p_i/\rho_ic^2$ and $\hat{\gamma}_i=\hat{\gamma}(\Theta_i)$, we will use here the Taub-Matthews equation of state\citep{Ryu+06,Mignone+2007},
\begin{eqnarray}\nonumber
h(\Theta)&=&\frac {5}{2}\Theta+\sqrt{1+\frac{9}{4}\Theta^2}\ ,
\\ \label{eq:EoS-TM}
\hat{\gamma}(\Theta)&=&\frac{h(\Theta)-1}{h(\Theta)-1-\Theta}=\frac{1}{6}\left(8-3\Theta+\sqrt{4+9\Theta^2}\right)\ ,\quad
\\ \nonumber
\Theta &=& \frac{(5-3\hat{\gamma})(\hat{\gamma}-1)}{3\hat{\gamma}-4} = \frac{1}{8}\left(5h-\sqrt{16+9h^2}\right)\ .
\end{eqnarray}
Denoting the upstream and downstream regions of a 1D shock by $i$ and $j$ respectively, the mass and energy equations in Table~\ref{table:1D} imply
\begin{equation}
\frac{e_j}{\rho_j} = \Gamma_{ij}\frac{w_i}{\rho_i}-\frac{p_i}{\rho_j}\quad\Longleftrightarrow\quad 
1+\frac{\Theta_j}{\hat{\gamma}_j-1}=\Gamma_{ij}h_i-\Theta_i\frac{\rho_i}{\rho_j}\ .
\end{equation}
We consider shock 1 for which $i=0$ and $j=1$, where for a strong shock ($\Theta_0\ll\Theta_1\Leftrightarrow h_0-1\ll h_1-1$) this reduces in the downstream region to $\Theta_1=(\hat{\gamma}_1-1)(\Gamma_1h_0-1)$, which according to Eq.~(\ref{eq:EoS-TM}) implies
\begin{equation}\label{eq:gamma1a}
\hat{\gamma}_j = \frac{4\Gamma_{ij}h_i+1}{3\Gamma_{ij}h_i}\ \ \longrightarrow\ \ 
\hat{\gamma}_1 = \frac{4\Gamma_{1}h_0+1}{3\Gamma_{1}h_0}\;,
\end{equation}
such that the conditions in region 1 are given by \citep{blandford1976fluid}
, which simplify further using Eq.~(\ref{eq:gamma1a}),
\begin{eqnarray}\nonumber
\frac{\rho_1}{\rho_0} &=& \frac{\Gamma_1[(4\Gamma_1+3)h_0+1]}{\Gamma_1h_0+1}\ ,
\\
\frac{p_1}{\rho_0c^2} &=& \frac{[(4\Gamma_1+3)h_0+1](h_0\Gamma_1-1)}{3h_0}\ ,\quad
\\ \nonumber
\frac{w_1}{\rho_0c^2} &=& \frac{[(4\Gamma_1+3)h_0+1](4\Gamma_1^2h_0^2-1)}{3h_0(\Gamma_1h_0+1)}\ ,
\\ \nonumber
u_{s1}^2 &=&\Gamma_{s1}^2-1 = \frac{[(4\Gamma_1+3)h_0+1]^2\Gamma_1^2(\Gamma_1-1)}{1+\Gamma_1[2h_0(1+\Gamma_1[h_0(4\Gamma_1+5)-1])-1]}\ .
\end{eqnarray}
The self-consistency strong shock condition reads
\begin{equation}
1\ll\frac{h_1-1}{h_0-1}=\hat{\gamma}_1\frac{\Gamma_1h_0-1}{h_0-1}=\hat{\gamma}_1\left[1+\frac{h_0(\Gamma_1-1)}{h_0-1}\right]\ ,
\end{equation}
and is always satisfied for a relativistic $\Gamma_1\gg1$, while for $\Gamma_1-1\lesssim1$ it corresponds to 
$\beta_1^2\gg p_0/\rho_0c^2$, which in all cases corresponds to a large Mach number
$\mathcal{M}_1=u_{s1}/u_{c_s,0}\gg1$ where $u_{c_s,0}=(\beta_{c_s,0}^{-2}-1)^{-1/2}$ 
is the proper sound speed in region 0 and $c_{s,0}=\beta_{c_s,0}\,c$ is the dimensionless sound speed.

Making a further assumption of an upstream that is not relativistically hot ($\Theta_0\ll1\Leftrightarrow h_0-1\ll1$), the expression for the adiabatic index reduces to
\begin{equation}\label{eq:gamma1}
\hat{\gamma}_j = \frac{4\Gamma_{ij}+1}{3\Gamma_{ij}}\ \ \longrightarrow\ \ 
\hat{\gamma}_1 = \frac{4\Gamma_{1}+1}{3\Gamma_{1}}\;.
\end{equation}

For demonstration purposes we will consider a cold region 0 where $p_0=0$, $e_0=w_0=\rho_0c^2$ and $h_0=1$, for which the shock is always strong, and the conditions in region 1 are given by
\begin{eqnarray}\nonumber\label{eq:region1_cold0}
\rho_1 &=& 4\Gamma_1\rho_0\ ,\quad
p_1 = \frac{4}{3}u_1^2\rho_0c^2\ ,\quad
e_1=4\Gamma_1^2\rho_0c^2\ ,\quad
\\
e_{\rm int,1} &=& \frac{4\Gamma_1u_1^2}{\Gamma_1+1}\rho_0c^2\ ,\quad 
w_1 = 4\Gamma_1^2\left(1+\frac{\beta_1^2}{3}\right)\rho_0c^2\ ,
\\ \nonumber
\beta_{s1} &=& \frac{4\Gamma_1u_1}{4\Gamma_1^2-1}\ ,\quad
u_{s1} = \frac{4\Gamma_1 u_1}{\sqrt{8\Gamma_1^2+1}}\ ,\quad
\Gamma_{s1} = \frac{4\Gamma_1^2-1}{\sqrt{8\Gamma_1^2+1}}\ .
\end{eqnarray}
Similarly
\begin{eqnarray}
u_1 &=& \frac{1}{2}\sqrt{u_{s1}^2-2+\sqrt{4+5u_{s1}^2+u_{s1}^4}}\ ,
\\ \label{eq:1Dbeta_sh}
\beta_{1,s1} &=& \frac{\beta_{s1}-\beta_1}{1-\beta_1\beta_{s1}} = \frac{\beta_1}{3}\ ,
\end{eqnarray}
where Eq.~(\ref{eq:1Dbeta_sh}) means that for our equation of state, in the rest frame of the downstream fluid (region 1), the speed at which the shock 1 is receding is a third of the incoming upstream speed.

\begin{figure}
	\centering
	\includegraphics[trim={0cm 0cm 0cm 0cm},clip,scale=0.6,width=0.48\textwidth,
 height=0.44\textheight
]{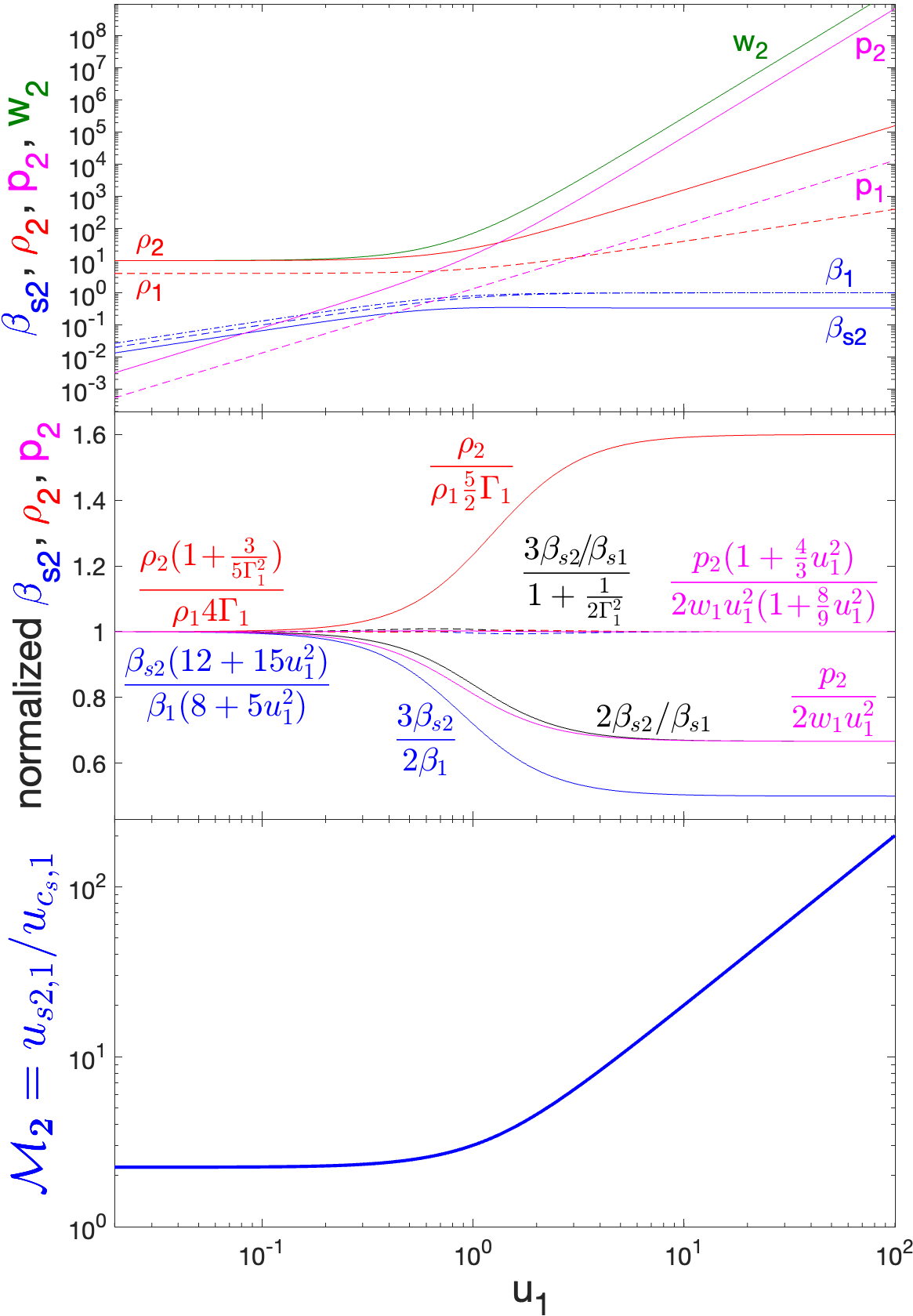}
	\vspace{-0.5cm}
	\caption{Results for shock reflection in 1D ($\alpha_1\to0$). All, where all velocities are measure in referece frame $S$ -- the rest frame of the reflecting wall, the unshocked fluid (region 0; assumed to be cold, $p_0=0$) and the doubly shocked fluid (region 2), while the singly shocked fluid (region 1) moves at a dimensionless velocity $\beta_1$, proper velocity $u_1$ and Lorentz factor $\Gamma_1$. \emph{\textbf{Top}}: The reflected shock velocity, $\beta_{s2}$, the pressure ($p$, in units of $\rho_0c^2$) and proper rest-mass density ($\rho$, in units of $\rho_0$) in regions 1 and 2, as well as $\beta_1$ and enthalpy density of region 2, $w_2$ (in units of $\rho_0c^2$), all shown as a function of $u_1$. \emph{\textbf{Middle}}: analytic approximations for $\beta_{s2}$, $\rho_2$ and $p_2$; the solid lines are simpler approximations that hold in the Newtonian limit ($u_1\ll1$), while the dashed lines are more refined approximations that hold for all $u_1$ values. \emph{\textbf{Bottom}}: The Mach number of the reflected shock, $\mathcal{M}_2=u_{s2,1}/u_{c_s,1}$, defined as the ratio of the proper velocity of the reflected shock ($s2$) relative to the upstream (region 1), $u_{s2,1}$, and the proper sound speed in the upstream, 
 $u_{c_s,1}=\beta_{c_s,1}(1-\beta_{c_s,1}^{2})^{-1/2}$ 
where $c_{s,1}=\beta_{c_s,1}\,c$ is the sound speed in the fluid rest frame.}
	\label{fig:ref1D}
\end{figure}

In the second step,
the set of 4 equations in \S\ref{sec:2D-fixed-volume} or \S\ref{sec:2D-fixed-fluid} for the mass (Eq.~(\ref{eq:SR-M1}) or (\ref{eq:SR-M2})), energy  (Eq.~(\ref{eq:SR-E1}) or (\ref{eq:SR-E2})), $x$-momentum  (Eq.~(\ref{eq:SR-Px1}) or (\ref{eq:SR-Px2})), and  $y$-momentum  (Eq.~(\ref{eq:SR-Py1}) or (\ref{eq:SR-Py2})), together with the equation of state (Eqs.~(\ref{eq:EoS-gamma}), (\ref{eq:EoS-TM})) provide 5 equations for the 5 unknowns ($e_2$, $p_2$, $\rho_2$, $v_2=\beta_2c$, $\alpha_2$), which can be solved in terms of the 5 knowns ($e_1$, $p_1$, $\rho_1$, $v_1=\beta_1c$, $\alpha_1$). We have not counted here $\beta_{s1}$ that can be calculated in the first step or $\beta_{s2}=\beta_{s1}\sin\alpha_2/\sin\alpha_1$.

\subsection{The one-dimensional case}
\label{sec:sol_1D}

In the limit $\alpha_1\to0$ the shock reflection reduces to a 1D problem and one can use the corresponding equations from Table~\ref{table:1D}. The results for this case are shown in Fig.~\ref{fig:ref1D}. In the Newtonian regime the reflected shock is rather weak, with a Mach number $\mathcal{M}_2\to\sqrt{5}\approx2.236$, while in the relativistic regime it is strong, with a Mach number $\mathcal{M}_2(u_1\gg1)\approx2u_1$ (see \textit{lower panel} of Fig.~\ref{fig:ref1D}). For any fixed $u_1$, the limit $\alpha_1\to0$ is always a super-luminal case (according to equation (\ref{eq:vp-reflection})).
Below we show that in the super-luminal case only the weak shock RR solution exists. Hence, the one-dimensional shock reflection always corresponds to the weak shock RR solution.

\subsection{The two-dimensional parameter space}
\label{sec:sol_2D}

Since the parameter space of the shock reflection problem is generally five-dimensional (see the end of \S\ref{sec:solutions}), we choose to focus on the case where the unshocked region~0 is cold. The properties of the compressible fluid (e.g. gas or plasma) in the singly shocked region~1 are then uniquely determined by $u_1$ (according to Eqs.~(\ref{eq:region1_cold0})). This, in turn, causes the parameter space to become two dimensional -- each point is fully specified by the proper velocity $u_1$ and the incidence angle $\alpha_1$.

\begin{figure}
\centering
\includegraphics[width=0.99\columnwidth]{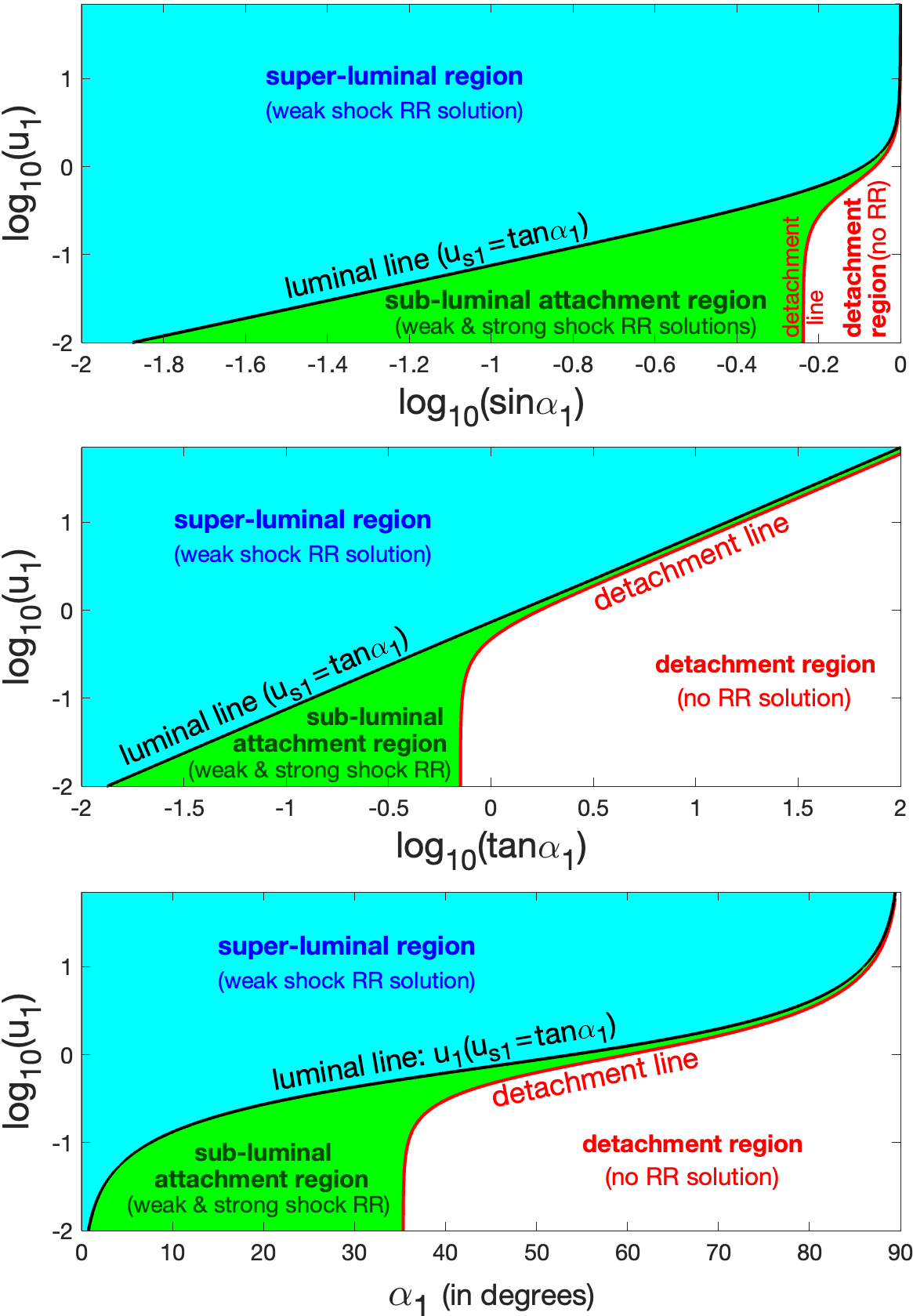}
\caption{The different regions and critical lines in the $u_1$\,--\,$\alpha_1$ parameter space, shown in terms of $\log_{10}(u_1)$ versus $\log_{10}(\sin\alpha_1)$ (\textbf{\textit{top panel}}), $\log_{10}(\tan\alpha_1)$ (\textbf{\textit{middle panel}}), and $\alpha_1$ (\textbf{\textit{bottom panel}}). The luminal line (\textit{in black}; $u_{s1}=\tan\alpha_1$) separates the super-luminal region (\textit{\cyan{cyan shading}}) and the sub-luminal, attachment region (\textit{\green{green shading}}). The latter is separated by the detachment line (\textit{\red{in red}}) from the detachment region (\textit{in wight}, where there is no RR solution). 
}	
\label{fig:critical_lines}
\end{figure}

Upon solving the equations, we find that the parameter space divides into different regions according to the nature and number of the solutions. Figure~\ref{fig:critical_lines} shows the different regions in the $u_1$\,--\,$\alpha_1$ parameter space, and the critical lines that separate between them. This is displayed by showing $\log_{10}(u_1)$ in the $y$-axis versus $\log_{10}(\sin\alpha_1)$ (\textit{top panel}), $\log_{10}(\tan\alpha_1)$ (\textit{middle panel}), and $\alpha_1$ (\textit{bottom panel}) in the $x$-axis. 

The luminal line (\textit{in black}) is defined as the line in the parameter space, for which the collision point P moves at the speed of light, $v_p=c$.  Equivalently, it also corresponds to $u_{s1}=\tan\alpha_1$. 
This line separates between the super-luminal region (in \textit{\cyan{cyan shading}}) and the sub-luminal regions beneath it. According to equation~(\ref{eq:region1_cold0}), the luminal relationship reads 
\begin{equation}
\frac{u_1} {\tan\alpha_1}=\frac{\sqrt{8\Gamma_1^2+1}}{4\Gamma_1u_1}\approx 
\begin{cases}
\frac{3}{4}=0.75  &\textrm{for}\ u_1\ll1\ ,\\
\frac{1}{\sqrt{2}}\approx0.7071  &\textrm{for}\ u_1\gg1\ ,\quad\quad\quad\quad
\end{cases}
\end{equation} 
such that $u_1/\tan\alpha_1$ varies by only 6\% throughout.

In the sub-luminal region (below the luminal line), there 
exists a region where no solutions to the RR 
equations exist (\textit{in white}). Therefore, in this 
region, RR is not possible, and instead 
there is MR/IR. The critical line that bounds the region 
with RR solutions is termed the detachment line (\textit{\red{in red}}), as is customary in Newtonian 
shock reflection terminology \citep{benDor1987}. The region with no RR solutions (\textit{in white}) is therefore termed the detachment region, as in this region the incident shock is detached from the reflecting wall. The regions with RR are called the attachemnt regions since there the incident shock is attached to the wall (at point P). The region between the luminal line and the detachment line (in \textit{\green{green shading}}) is termed the sub-luminal attachment region.

\begin{figure}
	\centering
	\includegraphics[trim={6cm 5cm 1cm 8.5cm},clip,scale=0.6,width=0.65\textwidth]{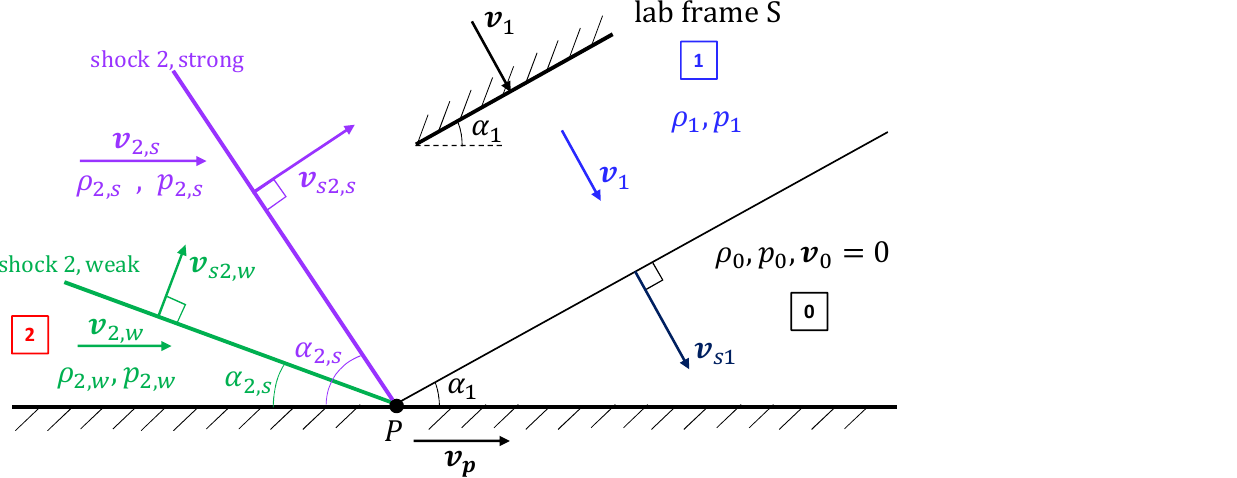}
	\\ \vspace{1.0cm}
	\includegraphics[trim={6cm 5cm 1cm 10.6cm},clip,scale=0.6,width=0.65\textwidth]{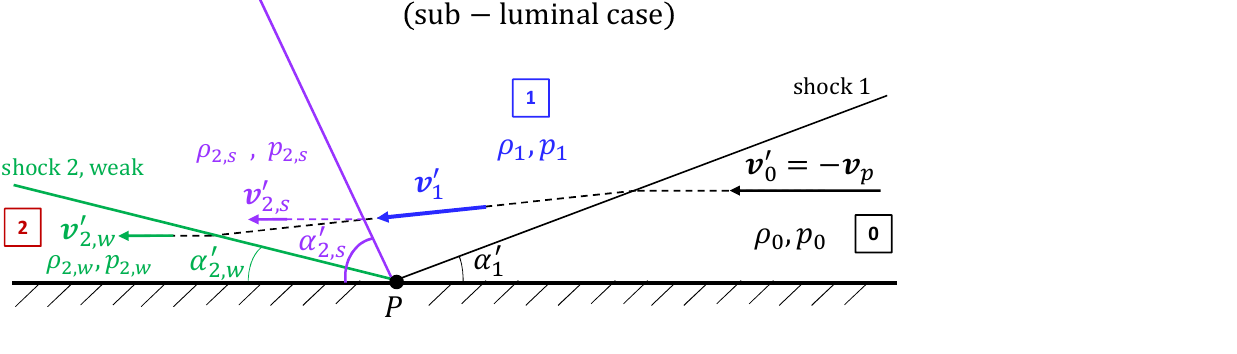}
	\vspace{-0.5cm}
	\caption{\textit{\textbf{Top}}: The shock reflection problem in the rest frame $S$ of the unshocked fluid (region 0). 
\textit{\textbf{Bottom}}:
In the sub-luminal case ($v_p<c$), one can conveniently transform to the frame $S'$ where point $P$ is at rest and the flow is steady and easier to solve.
}
	\label{fig:reflection_strong_and_weak}
\end{figure}

In the sub-luminal attachment region, we encounter (in \S\,\ref{sec:sol_2DS}) two solutions to the equations, which we classify as the ``weak'' and ``strong'' solutions, based on the pressure values in the doubly-shocked region ($p_2$), which reflect the strength of the reflected shock. These solutions are depicted schematically in Figure~\ref{fig:reflection_strong_and_weak}, and their properties are explored in the subsequent sections. However, we find that
the weak shock solution smoothly crosses the luminal line into the super-luminal region, while the strong shock solution no longer exists in the super-luminal region.

In the strong shock RR solution, the doubly shocked region 2 is always subsonic in frame $S'$, implying that it is in causal contact with point P. This means that it can act on point P and exert a force on it with a component along the normal to the wall. This can either cause it to detach from the wall (in the dual region discussed in \S\,\ref{sec:dual-region}), or to transition to the weak shock RR solution (as demonstrated, e.g., in \citep{Bera+23}. Indeed, the strong shock RR solution is theoretically known to be unstable\cite{hornung1997technical}, and was experimentally observed only under special conditions\cite{Ben-Dor92,sivaprasad2023strong}. In the weak shock RR solution, on the other hand, the doubly shocked region 2 is supersonic in frame $S'$ throughout almost all of the parameter space where it exists, implying that it is not in causal contact with point P and can therefore not cause it to detach from the wall. This also causes the weak sock RR solution to be more stable.

The critical line, where $\beta'_{2,w}=c_{s,2,w}$ (where the subscript `w' stands for the weak shock RR solution), is the sonic line for the weak shock RR solution, which is the analog of the sonic boundary in Newtonian shock reflection studies \citep{Ben-Dor92,hryniewicki2017transition}. We find that it is very close to the detachment line (for a fixed $u_1$ value its $\tan\alpha_1$ value is typically lower by several tenths of a percent compared to that of the detachment line). Similar findings were observed in the context of Newtonian shock reflection\citep{Ben-Dor92}.
Due to the close proximity of the two lines, it is very challenging experimentally to distinguish between them. Certain experiments conducted in the Newtonian regime have suggested that the boundary of the RR region corresponds to the sonic line rather than the detachment line\citep{lock1989experimental}, while certain numerical simulations have indicated that this boundary extends slightly to the right of the detachment line\citep{srikumar2018numerical}. 

The sonic line always lies in the sub-luminal region. This arises since it corresponds to $\beta_p=(\beta_2+c_{s2})/(1+\beta_2c_{s2})<1$, i.e. the sonic condition implies that $v_p$ is equal to the lab-frame speed of a sound wave moving in region 2 parallel to the wall, which must therefore be less than $c$.
As the sonic line almost coincides with the detachment line, it effectively separates between the detachement (or weak solution subsonic) region (\textit{in white}), where there is no RR (but only MR/IR) and the attachement regions -- the sub-luminal detached region (in \textit{\green{shaded green}}) and the super-luminal region (in \textit{\cyan{shaded cyan}}). 

\begin{figure}
	\centering
	\includegraphics[trim={0cm 0cm 0cm 0cm},clip,scale=0.6,width=0.48\textwidth]{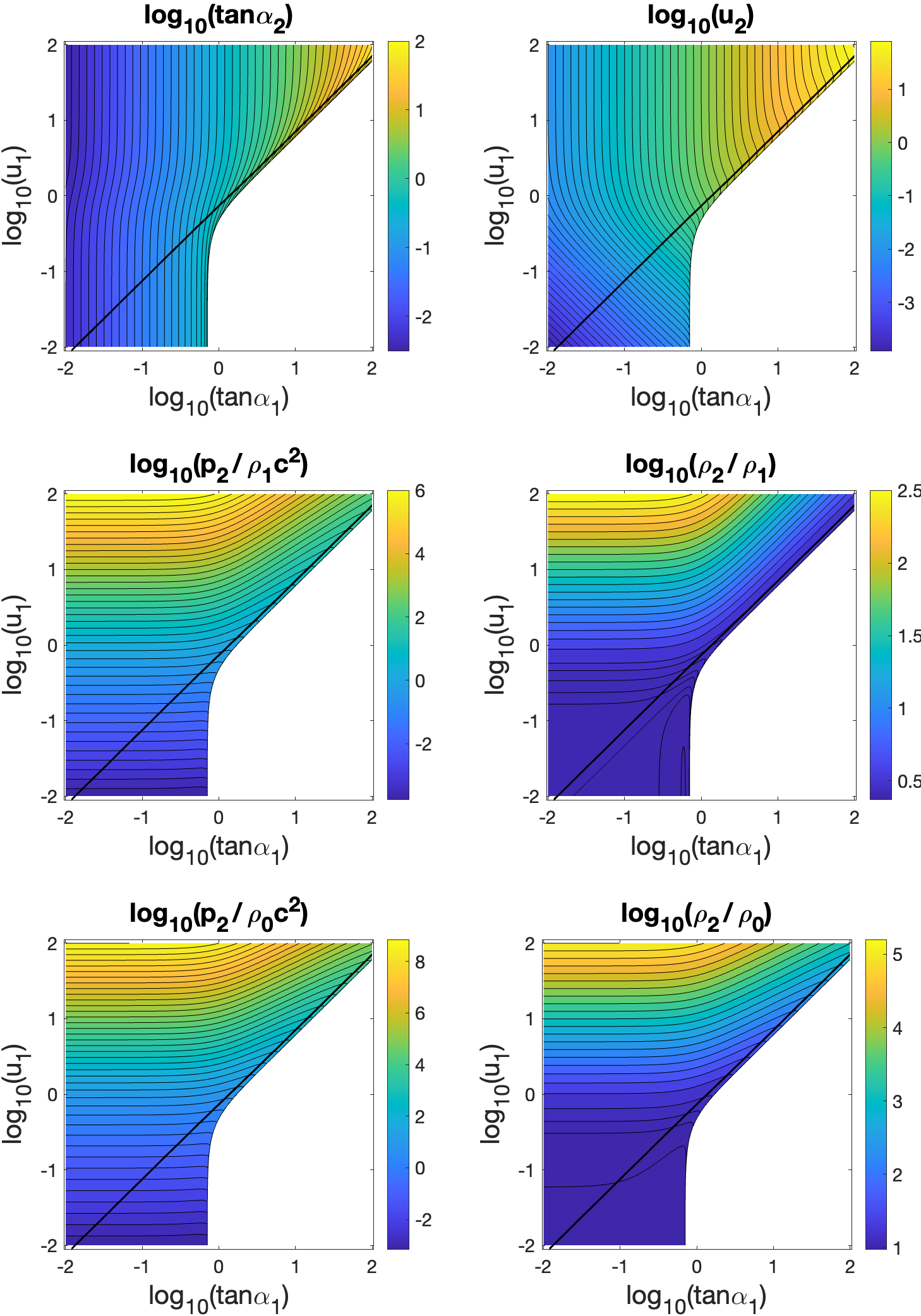}
	\caption{Results for shock reflection in 2D, for the weak shock RR solution, shown as a function of the proper velocity of the singly shocked fluid (region 1), $u_1$, and the incidence angle $\alpha_1$ (both measured in the rest frame of the reflecting wall), in the $\log_{10}(u_1)$\,--\,$\log_{10}(\tan\alpha_1)$ plane. We show contour lines for $\log_{10}(\tan\alpha_2)$ where $\alpha_2$ is the angle of the reflected shock front relative to the wall (\emph{top-left panel}), as well as the hydrodynamic variables in region 2, containing the doubly-shocked fluid: its proper speed ($u_2$; \emph{top-right panel}), its pressure ($p_2$) normalized by $\rho_1 c^2$ (\emph{middle-left panel}) or by $\rho_0 c^2$ (\emph{bottom-left panel}) and its proper rest-mass density ($\rho_2$) normalized by that or regions 1 ($\rho_1$; \emph{middle-right panel}) or 0 ($\rho_0$; \emph{bottom-right panel}). In each panel  the luminal line ($v_p=c$) is shown in black, while the white region in the bottom-right is the subsonic region where there is no regular shock reflection (see Fig.~\ref{fig:critical_lines}).
 }
	\label{fig:2Dresults_tan}
\end{figure}

\begin{figure}
	\centering
	\includegraphics[trim={0cm 0cm 0cm 0cm},clip,scale=0.6,width=0.47\textwidth]{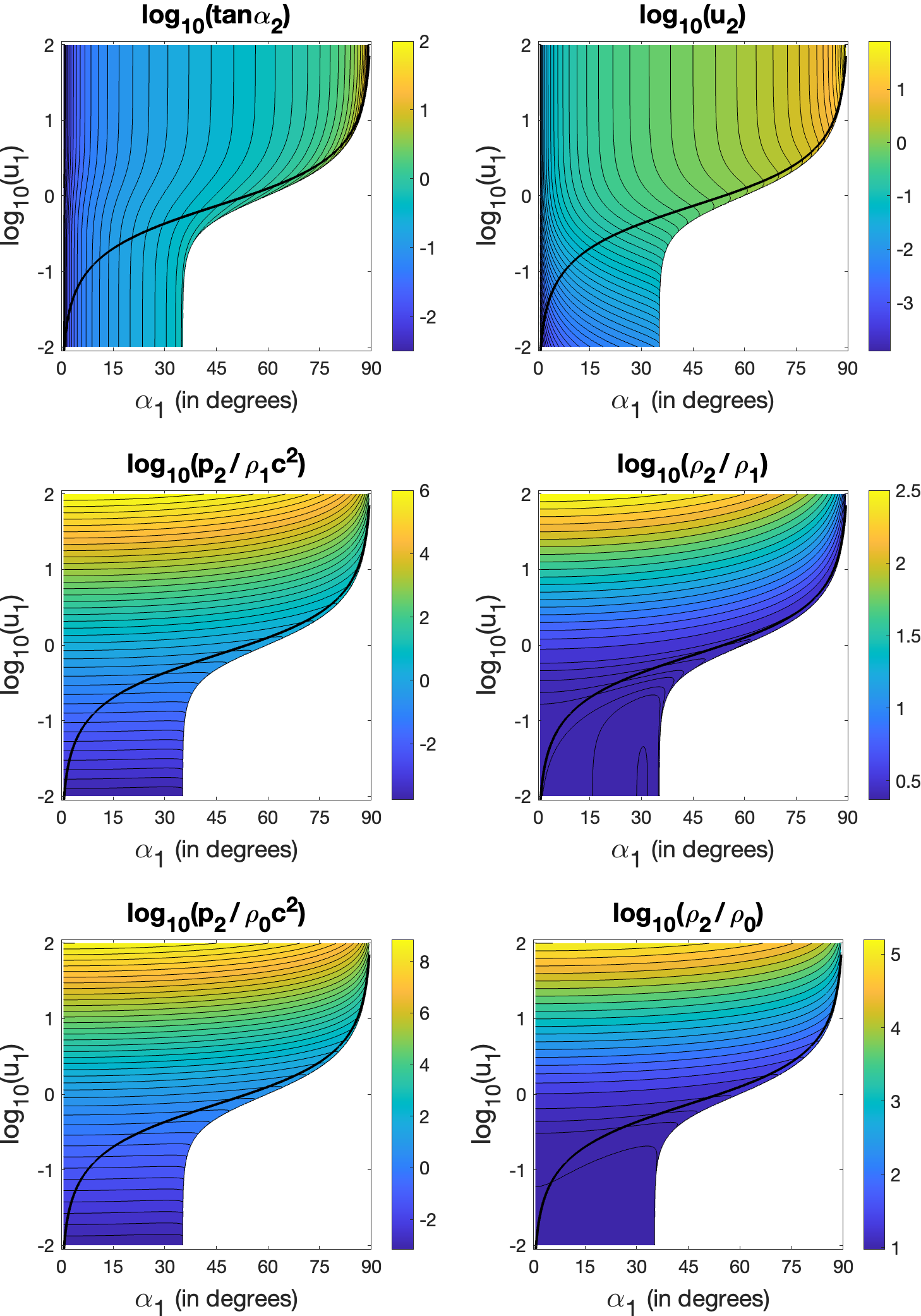}
	\caption{The same as Fig.~\ref{fig:2Dresults_tan}, but in the $\log_{10}(u_1)$\,--\,$\alpha_1$ plane.}
	\label{fig:2Dresults_alpha}
\end{figure}

\begin{figure}
   \centering
   \includegraphics[trim={0cm 0cm 0cm 0cm},clip,scale=0.6,width=0.47\textwidth]{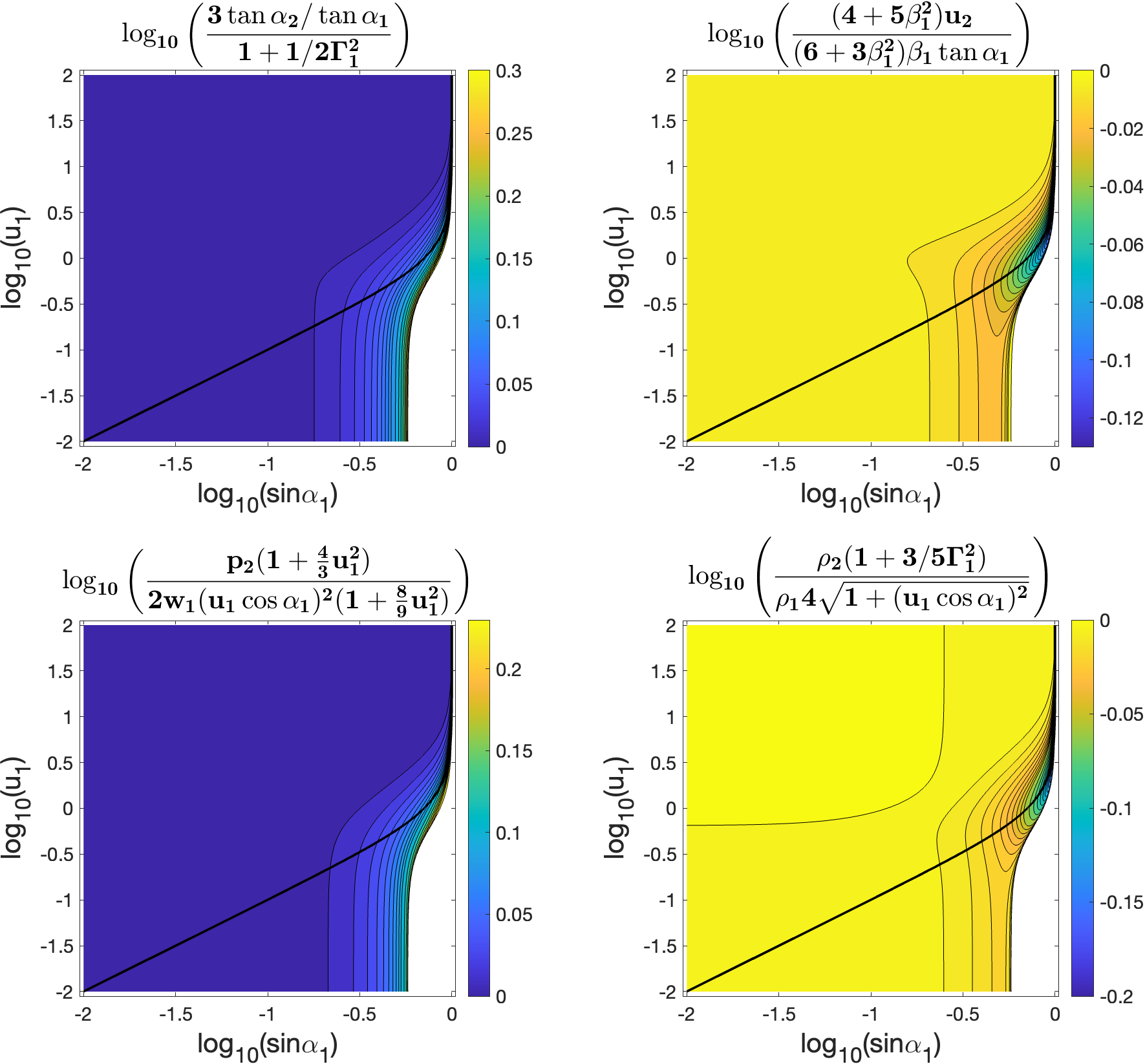}
   \caption{Results in the $\log_{10}(u_1)$\,--\,$\log_{10}(\sin\alpha_1)$ plane for the weak shock RR solution of a 2D shock reflection, where each quantity is normalized by an analytic function (written in the title of each panel) that captures most of its variation, with rather small deviations throughout most of the parameter space.}
\label{fig:2Dresults_normalizad}
\end{figure}

We note that under our assumption of a cold unshocked region~0, in the Newtonian regime ($u_1\ll1$) the detachment and sonic lines both approach constant values, which are close to each other, in agreement with the well known Newtonian results. On the other hand, the luminal line corresponds to extremely small incidence angles in this regime, $\alpha_1\approx\sin\alpha_1=\beta_{s1}\approx\frac{4}{3}\beta_1\ll1$, which is essentially why it could have been ignored so far. In the relativistic regime, however, the limunal line corresponds to $\tan\alpha_1=u_{s1}\approx\sqrt{2}\,\Gamma_1\gg1$, such that the super-luminal region occupies most of the relevant parameter space, and must be taken into account. Moreover, in the relativistic regime the detachment and sonic lines are quite close to the luminal line, their $\tan\alpha_1$ for the same $u_1$ being larger by only about 18\%. In the relativistic regime all of these three critical lines correspond to incidence angle very close to $90^\circ$.

\subsection{The two-dimensional weak shock solution}
\label{sec:sol_2DW}

The 2D results for the weak shock RR solution are shown in Figs.~\ref{fig:2Dresults_tan}, \ref{fig:2Dresults_alpha}  and \ref{fig:2Dresults_normalizad}. Figure~\ref{fig:2Dresults_tan} shows results in the $\log_{10}(u_1)$\,--\,$\log_{10}(\tan\alpha_1)$ plane while Fig.~\ref{fig:2Dresults_alpha} shows the same results in the $\log_{10}(u_1)$\,--\,$\alpha_1$ plane. Results are shown for $\tan\alpha_2$ where $\alpha_2$ is the angle between the reflected shock front and the reflecting wall  (\emph{top-left panel}), as well as the hydrodynamic variables in region 2 (containing the doubly-shocked fluid), namely its proper speed ($u_2$; \emph{top-right panel}), its pressure ($p_2$) normalized by $\rho_1 c^2$ (\emph{middle-left panel}) or by $\rho_0 c^2$ (\emph{bottom-left panel}) and its proper rest-mass density ($\rho_2$) normalized by that or regions 1 ($\rho_1$; \emph{middle-right panel}) or 0 ($\rho_0$; \emph{bottom-right panel}). In each panel  the luminal line ($v_p=c$) is shown in black, while the white region in the bottom-right is the detachment region where there is no RR solution (see Fig.~\ref{fig:critical_lines}).

In Fig.~\ref{fig:2Dresults_normalizad} the same quantities are normalized by a simple analytic function that captures most of its variation.
There are relatively small deviations from these analytic approximations (written in the title of each panel) throughout most of the parameter space. The deviations become larger in the sub-luminal case, below the black line that indicates $v_p=c$, and towards the detachment line beyond which there is no RR solution (but only MR/IR).
Figures.~\ref{fig:2Dresults_tan}, \ref{fig:2Dresults_alpha}  and \ref{fig:2Dresults_normalizad} clealy show that all the hydrodynamic variable describing in the doubly shocked region 2 smoothly transition across the luminal line.

For small incidence angles, $\alpha_1\ll1$, we use the expressions for the limiting case of a 1D normal shock reflection ($\alpha_1\to0$) from Fig.~\ref{fig:ref1D}. For $\tan\alpha_2$, in the $\alpha_1\ll1$ limit Eq.~(\ref{eq:vp-reflection}) implies  
\begin{equation}
\frac{\tan\alpha_2}{\tan\alpha_1} \approx \frac{\sin\alpha_2}{\tan\alpha_1}= \frac{\beta_{s2}}{\beta_{s1}} \approx \frac{1}{3}\left(1+\frac{1}{2\Gamma_1^2}\right)\ .
\end{equation}
For $p_2$ we use the 1D expression with a correction factor $\cos^2\alpha_1$ that corresponds to taking only the component of the momentum flux in region 1 in the direction normal to the wall. For the proper rest-mass density $\rho_2$ we use the 1D expression and using only the perpendicular component of the proper speed, $u_1\cos\alpha_1$, by replacing $\Gamma_1^2\to1+u_1^2\cos^2\alpha_1$. For $u_2$ there is no 1D analog, and we have $u_2\approx\frac{3}{2}\beta_1\tan\alpha_1$ in the Newtonian limit ($u_1\ll1$) and $u_2\approx\tan\alpha_2$ in the relativistic limit ($u_1\gg1$).

\subsection{The two-dimensional strong shock solution}
\label{sec:sol_2DS}

\begin{figure}
	\centering
	\includegraphics[trim={1.5cm 2.15cm 0cm 1.63cm},clip,scale=1,width=0.55\textwidth]{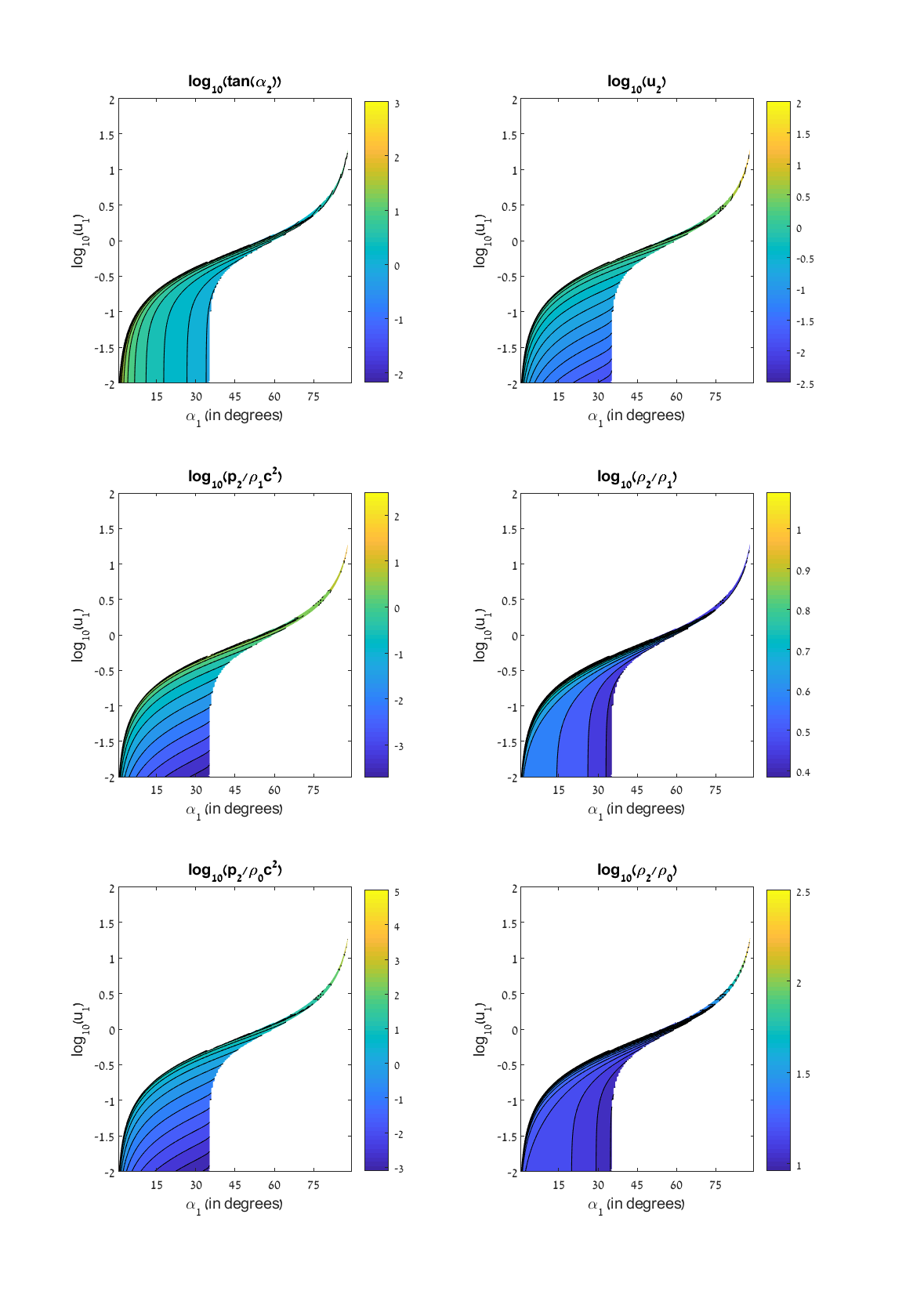}
	\vspace{-0.5cm}
	\caption{The same as Fig.~\ref{fig:2Dresults_alpha}, but for the strong shock RR solution. 
 }
\label{fig:2Dresults_alpha_strong}
\end{figure}

Fig.~\ref{fig:2Dresults_alpha_strong} shows the results for the strong shock RR solution of a 2D shock reflection. We find that the weak and strong shock solutions coincide along the detachment line. This arises since there they both correspond to the maximal deflection angle in frame S$^\prime$, beyond which there is no RR solution (in the detachment region). The two solutions grow apart away from the detachment line in the sub-luminal attchment region.

\begin{figure}
	\centering
	\includegraphics[trim={5cm 1.23cm 1cm 4.75cm},clip,scale=1,width=0.637\textwidth]{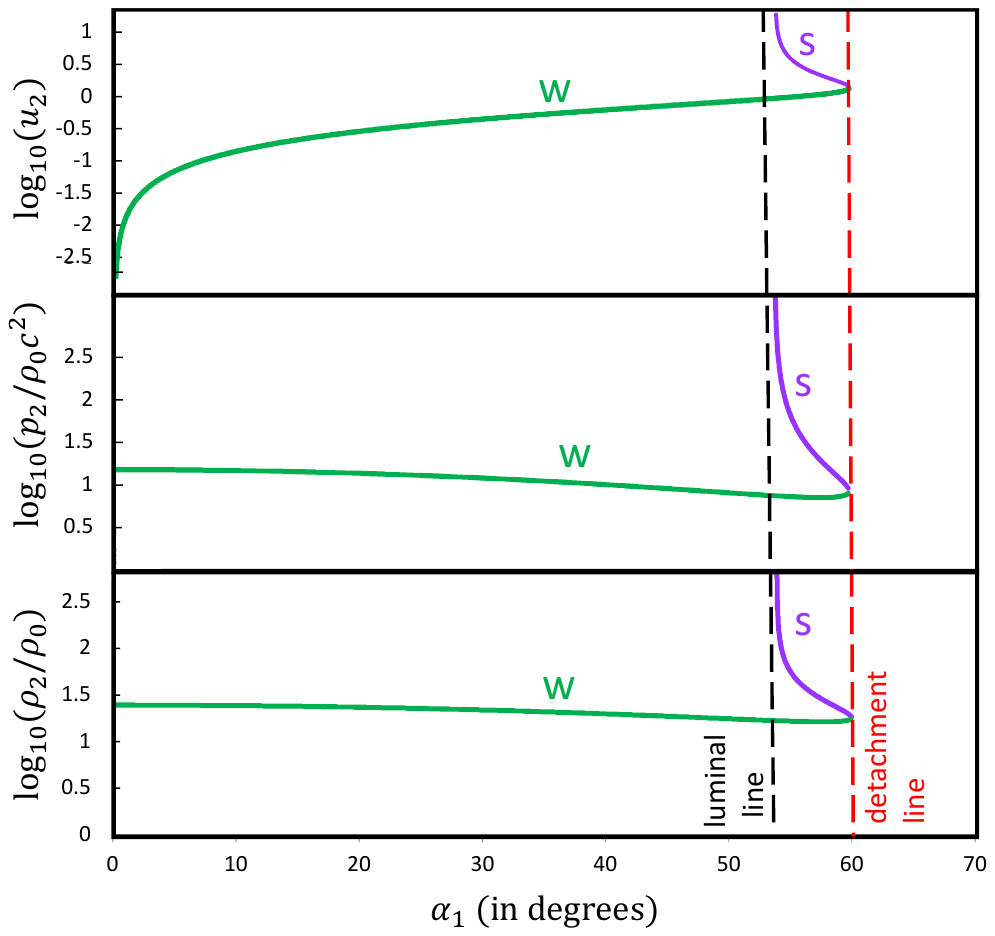}
	\vspace{-0.5cm}
	\caption{The proper speed, $u_2$, normalized pressure, $p_2/\rho_0c^2$, and normalized proper rest-mass density, $\rho_2/\rho_0$, in the doubly shocked region 2, for $u_1=1$ as a function of the shock incidence angle $\alpha_1$, for the weak (`W' in \textit{\darkgreen{green}}) and strong (`S' in \textit{\darkpurple{{purple}}}) shock RR solutions. The strong shock solution exists only between the luminal and detachment lines, while the weak shock solution smoothly transitions across the luminal line.
 }
\label{fig:bifurcation1}
\end{figure}

\begin{figure}
	\centering
	\includegraphics[trim={5cm 2.3cm 0cm 1.45cm},clip,scale=2,width=0.657\textwidth]{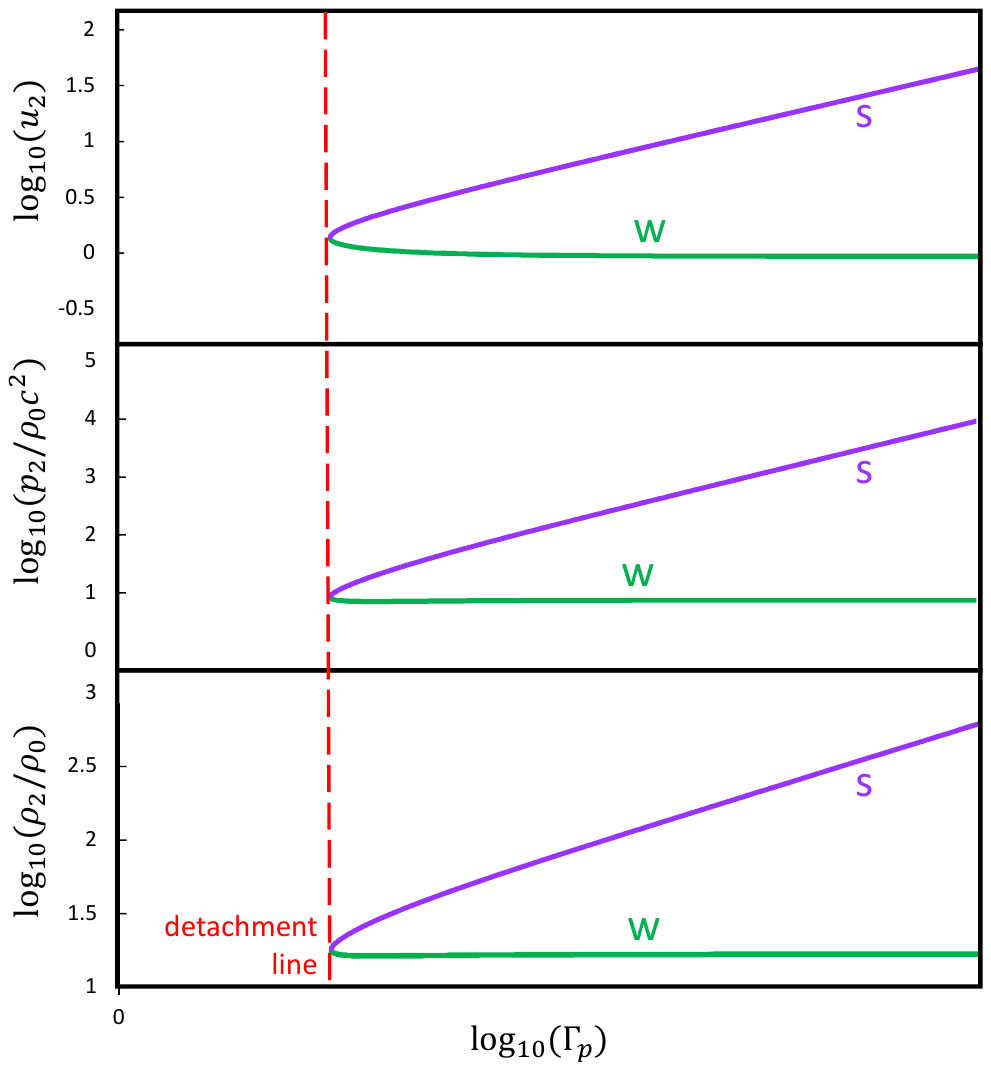}
	\vspace{-0.5cm}
	\caption{The same hydrodynamic variables in region 2 as in Fig.~\ref{fig:bifurcation1} but as a unction of $\Gamma_p$ (the Lorentz factor corresponding to the boost between the lab frame $S$ and the steady-state frame $S'$) in the sub-luminal attachment region between the detachment and luminal lines.
 }
\label{fig:bifurcation2}
\end{figure}

While the weak shock solution smoothly transitions across the luminal line, the strong shock solution displays a very different behavior in this respect. This very different behavior can conveniently be understood when considering what happens when the two solutions approach some point along the luminal line from the sub-luminal region.
This approach can conveniently be followed in frame S$^\prime$ considering a fixed $u_1$ (which corresponds to a fixed $\beta_{s1}$) while decreasing $\sin\alpha_1$ such that $\sin\alpha_1\to\beta_{s1}\Leftrightarrow\beta_p=\beta_{s1}/\sin\alpha_1\to1$. This is shown in Figures~\ref{fig:bifurcation1} and \ref{fig:bifurcation2} for $u_1=1$.

Figures~\ref{fig:bifurcation1} shows the (normalized) proper speed ($u_2$), pressure ($p_2$) and proper rest-mass density ($\rho_2$) in the doubly shocked region 2 as a function of the shock incidence angle $\alpha_1$, for a fixed $u_1=1$. The weak shock solution exists throughout the attachment region, at all incidence angles below the detachment angle (i.e. up to the detachment line), all hydrodynamic variables smoothly vary across the luminal line. On the other hand, the strong shock solution exists only between the luminal line and the detachment line, while $u_2$, $p_2$ and $\rho_2$ all diverge towards the luminal line.

Figure~\ref{fig:bifurcation2} shows the same hydrodynamic variable in the doubly shocked region 2 but as a unction of $\Gamma_p$ -- the Lorentz factor corresponding to the boost between the lab frame $S$ and the steady-state frame $S'$. The divergence of $u_2$, $p_2$ and $\rho_2$ atowards the luminal line (i.e. at large $\Gamma_p$ values) for the strong shock solution can clearly be seen here. Moreover, it can be seen that in this limit $u_2,\,\rho_2\propto\Gamma_p$ while $p_2\propto\Gamma_p^2$. 

This can be understood as follows. As $\Gamma_p\to\infty$ for a fixed $u_1$, $\tan\alpha_1\to u_{s1}(u_1)$ approaches a constant value, but in frame $S'$ we have $\tan\alpha'_1=\Gamma_p^{-1}\tan\alpha_1\to0$. For $\Gamma_p\gg1$ this implies $1\gg\alpha'_1\approx\tan\alpha'_1\approx u_{s1}(u_1)/\Gamma_p\propto\Gamma_p^{-1}$. Similarly, $\Gamma'_1=\Gamma_p\Gamma_1(1-\beta_p\beta_1\sin\alpha_1)\to\Gamma_p\Gamma_1(1-\beta_1\beta_{s1}(u_1))\propto\Gamma_p$. In the steady state frame $S'$ the first shock only very slightly deflects the fluid velocity (at an angle $\chi'<\alpha'_1\ll1$)
such that for the strong shock solution $\alpha'_2\approx\pi/2$ and the second shock must be almost perpendicular to achieve the same small deflection 
in the opposite direction. Such a nearly perpendicular relativistic shock (with $1\ll\Gamma'_1\propto\Gamma_p$) slows down the fluid at region 2 to a mildly 
relativistic velocity in frame $S'$ corresponding to $u_2\propto\Gamma_p$ in 
frame $S$. Similarly, its compression ratio scales as $\Gamma'_1\propto\Gamma_p$ explaining why $\rho_2\propto\Gamma_p$, while 
the downstream pressure scales as the upstream ram pressure, $p_2\propto (u'_1)^2\approx(\Gamma'_1)^2\propto\Gamma_p^2$.

The divergence of the hydrodynamic variable $u_2$, $p_2$ and $\rho_2$ in region 2 as the strong shock RR solution approaches the luminal line from the sub-luminal side, explains why this solution cannot reach or cross the luminal line.

\begin{figure}
	\centering
	\includegraphics[trim={0cm 0cm 0cm 0cm},clip,scale=2,width=0.48\textwidth]{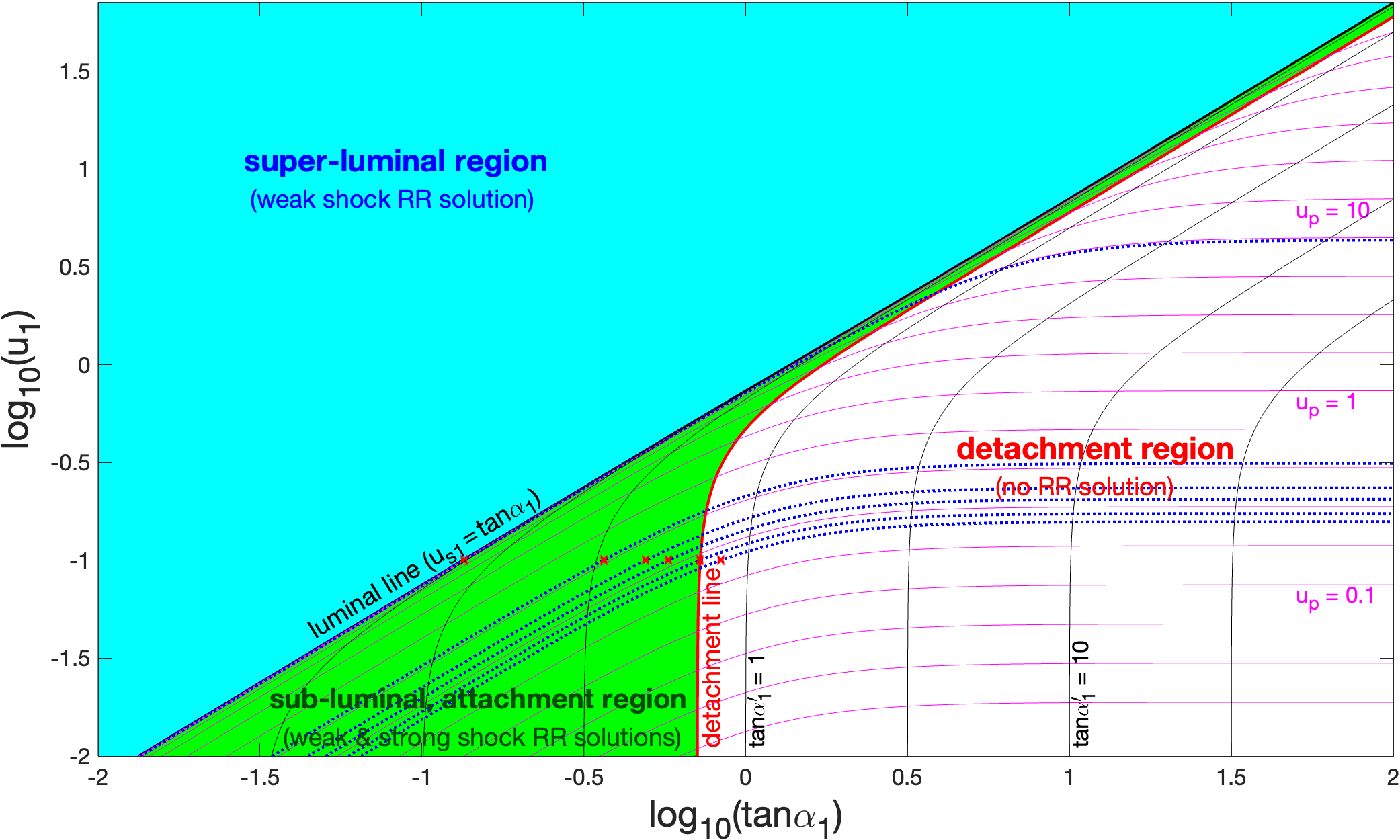}
	\vspace{-0.5cm}
	\caption{Constant $u_p$ (\magenta{magenta}; $\log_{10}(u_p)=-1.6,-1.4,...,2$) and constant $\alpha'_1$ (black; $\log_{10}(\tan\alpha'_1)=-1.5,-1,-0.5,0,0.5,1,1.5$) contours in the $\log_{10}(u_1)$\,--\,$\log_{10}(\tan\alpha_1)$ parameter space. They are both restricted to the sub-luminal region where frame $S'$ exists. The \textit{\blue{blue dotted}} lines and red \red{\textbf{x}} symbols correspond to the shock polars for shocks 1 and 2, respectively, that are shown in Fig.~\ref{fig_Shockpolars}.
 }
\label{fig:up_tan1p_contours}
\end{figure}

\subsection{Shock Polars in the sub-luminal region \& the Dual Region with both RR \& MR solutions}
\label{sec:dual-region}

\begin{figure*}
\includegraphics[trim={1.78cm 4.05cm 0.26cm 4.91cm},clip,width=0.98\textwidth]{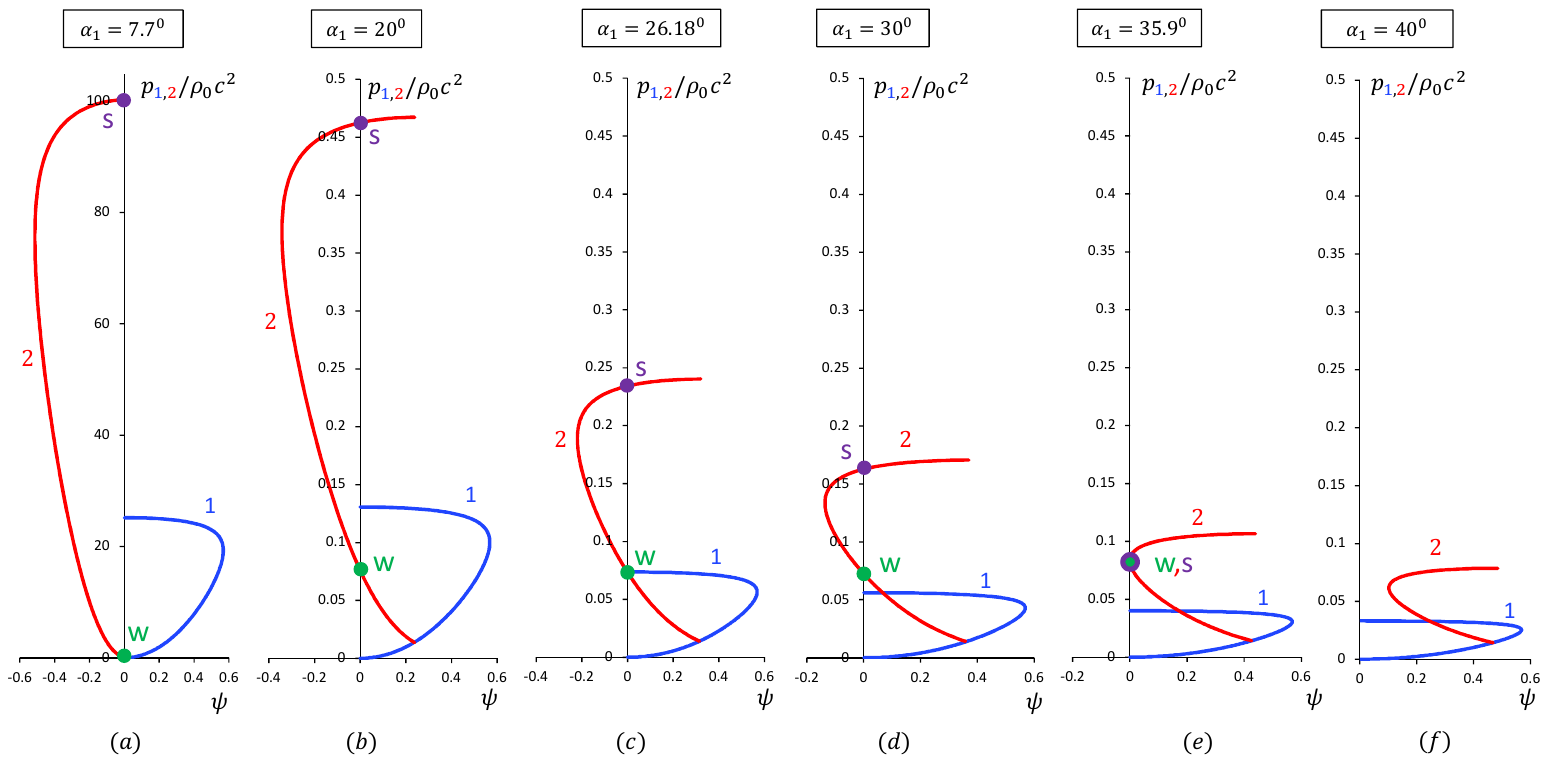}
    \caption{Shock polar diagrams for different incidence angles in the steady-state frame $S'$. The shock polars of the incident shock (\textit{\blue{blue line, 1}}) and of the reflected shock (\textit{\red{red line, 2}}) show the post-shock pressure $p_{\blue{1},\red{2}}$ vs. the deflection angle of the flow $\psi$ (relative to the flow direction in region 0). 
    The intersections of shock polar 2 with the $p$-axes are the weak shock (\textit{\darkgreen{green dot}}) and the strong shock (\textit{\darkpurple{purple dot}}) solutions. The polars of shock 2 are all for $u_1=0.1$ and: (a) $\alpha_1=7.7^0$, case of closed proximity to the luminal line. (b) $\alpha_1=20^0$, case with only RR. (c) $\alpha_1=26.18^0$, case of the mechanical-equilibrium criterion. (d) $\alpha_1=30^0$, case in the dual region. (e) $\alpha_1=35.9^0$, case of the detachment criterion. (f) $\alpha_1=40^0$, case of no RR solutions. The corresponding polars of shock 1 are for $u_p=6.12,\,0.419,\,0.314,\,0.274,\,0.231,\,0.210$.}
    \label{fig_Shockpolars}
\end{figure*}

Within the framework of Newtonian physics, the analysis of the shock reflection problem is typically carried out within the reference frame $S'$, where point P remains stationary. In this frame of reference, it is customary to describe the shock reflection using shock polar diagrams\citep{kawamura1956reflection}, represented by variables $(p,\chi')$, where $p$ is the pressure in the shocked region and $\chi'$ is the flow velocity deflection angle due to the shock. In this subsection, we employ these shock polar diagrams to illustrate the shift between various solutions identified in the sub-luminal regions. It is important to note that this analysis is applicable solely within the sub-luminal region of the parameter space, where a transformation to the frame $S'$ is possible.

The shock polar of the incident shock (shock 1) is the graphical representation of the relation between the post-shock pressure $p_1$ and the flow deflection angle $\psi=\chi'_1$ within the reference frame $S'$, for a fixed velocity of the flow in region 0, $\textit{\textbf{v}}'_0=-\textit{\textbf{v}}_p$, as in frame $S$ region 0 is at rest.
Hence, the shock polar for the incident shock (shock 1) is constructed by following constant-$u_p$ lines in the parameter space, which are depicted by the \textit{\magenta{magenta lines}} 
in Figure~\ref{fig:up_tan1p_contours}.

Similarly, the shock polar of the reflected shock (shock 2) graphically shows the relation between its post-shock pressure $p_2$ and the total flow velocity deflection angle $\psi=\chi'_1-\chi'_2$ (which is considered relative to the flow direction within region 0, which is parallel to the wall), for a fixed velocity of the flow in region 1, which corresponds to a single point ($u_1\,\alpha_1$) in the $u_1$\,--\,$\alpha_1$ parameter space. Given that the flow in region 1 has already experienced a deflection ($\chi'_1$; see Eq.~(\ref{eq:tan_chi_p})) and has a non-zero pressure ($p_1$; see Eq.~(\ref{eq:region1_cold0})), the shock polar of shock 2 commences with these specific pressure and deflection angle values, and is constructed by varying the possible angle $\alpha'_2+\chi'_1$ between $\textit{\textbf{v}}'_1$ and shock 2 in frame $S'$. When the total deflection angle along this second polar vanishes, $\psi=\chi'_1-\chi'_2=0$, the flow within region 2 adheres to the appropriate boundary condition along the wall. As a result, the point where the second shock polar intersects the $p$-axis signifies a solution to the shock reflection problem. The formalism involving steady-stare shock jump conditions in frame $S'$ that is used for calculating the shock polars is outlined in Appendix~\ref{sec:cons-steady}. We have verified that the solutions of the steady-state equations in frame $S'$ for the sub-luminal regime are identical to those from our formalism in frame $S$ that were derived in \S\,\ref{sec:lab_frame}.

Figure~\ref{fig_Shockpolars} shows various shock polar combinations. The polars of shock 2 are for
$u_1=0.1$ and different incidence angles $\alpha_1$ which are fixed for each polar (see the red \red{\textbf{x}} symbol in Fig.~\ref{fig:up_tan1p_contours}), while the corresponding polars of shock 1 are for $u_p=6.12,\,0.419,\,0.314,\,0.274,\,0.231,\,0.210$ (corresponding to the \textit{\blue{blue dotted}} lines in Fig.~\ref{fig:up_tan1p_contours}). As explained above, in the sub-luminal attachment region there are two solutions, which correspond to the two intersection points of shock polar 2 (corresponding to the reflected shock) with the $p$-axes. This is viewed in panels (a) through (d), for angles $\alpha_1<35.9^0$, and the two solutions are designated as `s' for strong and `w' for weak. Upon reaching the detachment/sonic line  at $\alpha_1=35.9^0$, the two solutions coincide (panel (e)), and for higher angles ($\alpha_1=40^0$ in panel (f)), there are no RR solutions available.

The case $\alpha_1=7.7^0$ (panel (a)) shows the shock polars for a case close to the luminal line ($u_p=6.12$; see the left red \red{\textbf{x}} symbol in Fig.~\ref{fig:up_tan1p_contours}). As previously discussed (see section  \S\ref{sec:sol_2DS}), the pressure of the strong shock solution becomes infinitely large as it approaches the luminal line. As is observed in the figure, the shock polar of shock 2 indeed exhibits a significant increase by about three orders of magnitude when compared to the other shock polar curves.

\begin{figure}
	\centering
	\includegraphics[trim={3.7cm 4.5cm 4.38cm 5.13cm},clip,width=0.48\textwidth]{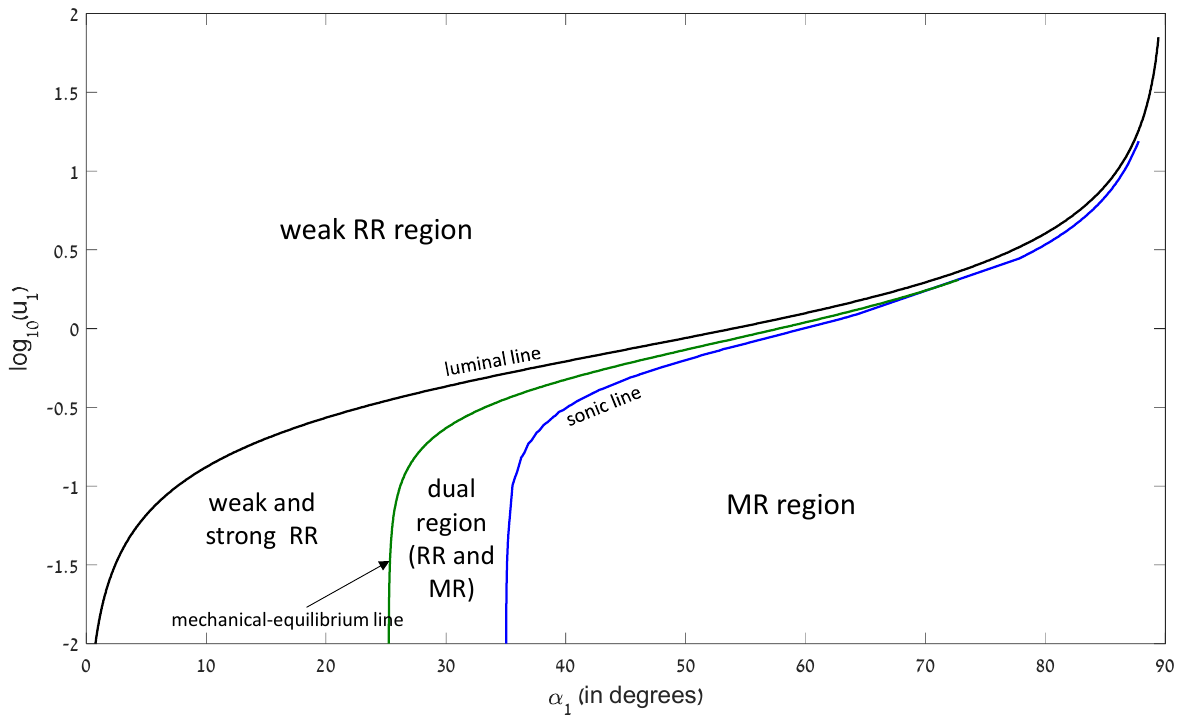}
   \caption{Critical lines in the $\log_{10}(u_1)$\,--\,$\alpha_1$ plane that bound regions with different shock reflection solutions. In addition to the luminal (\textit{black}) and detachment/sonic (\textit{\blue{blue}}) lines, we added here the mechanical equilibrium (\textit{\darkgreen{green}}) line, which bounds (to its bottom right) the region with MR, and the dual region between it and the detachment line where both RR and MR solutions exist.
 }
\label{fig:dual_region}
\end{figure}

Panel (c) illustrates a case in which the pressure of the weak solution matches the highest pressure point on the incident shock polar. This situation corresponds to the von Neumann mechanical equilibrium criterion, known in the Newtonian shock reflection problem\citep{vonNeuman1963,benDor1987}. According to this criterion, transitioning from RR to MR requires both reflection configurations to yield identical post-shock pressures. In the context of Newtonian physics, it has been observed that within the region bounded by the mechanical equilibrium line (established based on the mechanical equilibrium criterion) and the sonic line, both RR and MR can coexist simultaneously. Therefore, this intermediate region between these two lines is commonly referred to as the dual region. Figure \ref{fig:dual_region} illustrates the extension of the dual region to the relativistic case.

\section{Conclusions}
\label{sec:dis}

This work generalizes the classical problem of shock reflection, which has been studied in detail in the Newtonian regime, to the more general relativistic case, which is relevant in astrophysics. While in the Newtonian limit the problem can always be studied in a frame $S'$ where the flow is in steady state, which greatly simplifies the treatment, this is not always possible in the general case. In particular, we identify a new super-luminal regime where no such steady-state frame $S'$ exists. While for Newtonian shock this regime corresponds to extremely small incidence angles, in the relativistic regime it corresponds to almost all incidence angles (with the exception of those extremely close to $90^\circ$). Addressing the super-luminal regime required us to developed a new formalism in the lab frame $S$ where the unshocked region~0 is at rest, using the integral conservation laws. This formalism was applied to regular reflection (RR) for which it is most readily applicable, as in that case all regions are uniform and both the incident and reflected shock are planar, such that the integral conservation laws can be expressed as algebraic equations.

We have solved the resulting set of equations for the relatively simpler case of a cold unshocked medium, which reduces the relevant parameter space to two dimensions -- the proper speed of the singly shocked fluid ($u_1$) and the shock incidence angle ($\alpha_1$). This parameter spaced was mapped for the number and type of RR solutions, finding it divides into the following regions with corresponding types of solutions: 
\begin{enumerate}[leftmargin=0.4cm,labelwidth=0.4cm,itemsep=-0.0em]
\item \textbf{Super-luminal} region: \textbf{one} RR solution -- weak shock,
\item \textbf{Sub-luminal attachment} region: \textbf{two} RR solutions -- weak shock \& strong shock,
\item \textbf{Detachment} region: \textbf{no} RR solutions.
\end{enumerate}
In the detachment region only Mach or irregular reflectin (MR/IR) is possible. 
In addition, MR is also possible in part of the sub-luminal attachment region (which lies between the luminal and detachment lines)  -- the dual region that lies between the attachment line and the mechanical equilibrium line. In the super-luminal region there are no MR/IR solutions, such that only a single solution exists -- the weak shock RR solution.

Our results are related to the traditional shock polar description in the sub-luminal regions. In the super-luminal regime, however, such a description is no longer possible.

\begin{acknowledgments}
This research was funded in part by the ISF-NSFC joint research program under grant no. 3296/19 (J.G.). 
\end{acknowledgments}

\appendix

\section{Consistency with the steady-state analysis in the sub-luminal case}
\label{sec:cons-steady}

In the sub-luminal case ($v_p<c$) there is a reference frame $S'$ where point $P$ is at rest and the flow is steady. Therefore, a ruler at rest in $S'$ oriented parallel to the wall will Lorentz contract in frame $S$, $L_\parallel=L'_\parallel/\Gamma_p$, while in the perpendicular direction the length remains unchanged, $L_\perp=L'_\perp$. Hence, the angle of the shock front relative to the wall transforms as 
\begin{equation}
\tan\alpha_{i} = \frac{L_{i,\perp}}{L_{i,\parallel}} = \Gamma_p\frac{L'_{i,\perp}}{L'_{i,\parallel}} = \Gamma_p\tan\alpha'_{i}\ ,\quad(i=1,\,2)\ .
\end{equation}

Therefore, we can write the oblique shock jump conditions in this frame as
\begin{eqnarray}\nonumber
\rho_1u'_1\sin\alpha'_+ &=& \rho_2u'_2\sin\alpha'_2\ ,
\\ \nonumber
w_1(u'_1\sin\alpha'_+)^2+p_1 &=& w_2(u'_2\sin\alpha'_2)^2+p_2\ ,
\\ \label{eq:Sp-jump-conditions}
w_1\Gamma'_1u'_1\sin\alpha'_+ &=& w_2\Gamma'_2u'_2\sin\alpha'_2\ ,
\\ \nonumber
\beta'_1\cos\alpha'_+ &=& \beta'_2\cos\alpha'_2\ ,
\end{eqnarray}
where $\alpha'_+=\alpha'_2+\chi'$ and $\cos\chi'=-\hat{\beta}'_1\cdot\hat{y}'$ or
\begin{equation}\label{eq:tan_chi_p}
\tan\chi' = \frac{v'_{1x}}{v'_{1y}} = \frac{v_1\cos\alpha_1}{\Gamma_p(v_p-v_1\sin\alpha_1)}\ ,
\end{equation}
and $\tan\alpha'_2=\Gamma_p^{\,-1}\tan\alpha_2$ or
\begin{eqnarray}
\sin\alpha'_2 &=& \frac{\sin\alpha_2}{\sqrt{1+u_p^2\cos^2\alpha_2}} = \frac{\Gamma_{s2}}{\Gamma_p}\sin\alpha_2 = \frac{u_{s2}}{u_p}\ ,
\\
u_p^2&=&\Gamma_p^2-1=\frac{\beta_{s1}^2}{\sin^2\alpha_1-\beta_{s1}^2}=\frac{\beta_{s2}^2}{\sin^2\alpha_2-\beta_{s2}^2}\ .\quad
\end{eqnarray}
Together with the Lorentz transformationds of the velocities,
\begin{eqnarray}
\Gamma'_1=\Gamma_1\Gamma_p(1-\beta_1\beta_p\sin\alpha_1)\ ,\quad\Gamma'_2=\Gamma_1\Gamma_p(1-\beta_2\beta_p)\ ,\quad\ \   
\end{eqnarray}
it can be shown that Eqs.~(\ref{eq:Sp-jump-conditions}) are equivalent to Eqs.~(\ref{eq:SR-M1})\,--\,(\ref{eq:SR-Py1}) in frame $S$.

\end{document}